\documentclass{article}
\usepackage[utf8]{inputenc}
\usepackage{graphicx}     
\usepackage[numbers,sort&compress]{natbib}
\usepackage{float}      
\usepackage{placeins} 
\usepackage{indentfirst}  
\usepackage[letterpaper, margin=1in]{geometry}
\usepackage{amsmath}      
\usepackage{amsfonts}     
\usepackage{setspace}     
\usepackage{booktabs}
\usepackage{fancyhdr}     
\usepackage{lipsum}       
\usepackage{xcolor}

\begin{document}
\pagestyle{plain}
\raggedright


\vspace*{1cm}
\begin{center}
{\LARGE\textbf{Causal Inference with Missing Exposures and Missing Outcomes}} \\[1.5cm]
\end{center}
Kirsten E. Landsiedel$^{1}$, Rachel Abbott$^{2}$, Atukunda Mucunguzi$^{3}$, Florence Mwangwa$^{3}$, Elijah Kakande$^{3}$,
Edwin D. Charlebois$^{4}$, Carina Marquez$^{2}$, Moses R. Kamya$^{3,5,\dag}$, Laura B. Balzer$^{1,\dag,*}$ \\[1cm]

{\small
$^1$School of Public Health, University of California Berkeley, Berkeley, California, USA. \\
$^2$Division of HIV, Infectious Diseases and Global Medicine, University of California San Francisco, San Francisco, California, USA. \\
$^3$Infectious Diseases Research Collaboration, Kampala, Uganda.\\
$^4$Center for AIDS Prevention, University of California San Francisco, San Francisco, California, USA. \\
$^5$Department of Medicine, Makerere University, Kampala, Uganda. \\
}

$^\dag$Co-senior authors; $^*$Corresponding author \\[0.5cm]

\vspace{0.5cm}

\textbf{Corresponding Author:} \\
Laura B. Balzer \\
Mail:   2121 Berkeley Way West, Berkeley, CA 94720, USA \\
Email: laura.balzer@berkeley.edu \\[0.5cm]

\textbf{Source of Funding:} This work was supported, in part, by The National Institutes of Health (awards: R01AI151209 (CM), K23AI118592 (CM), U01AI099959, and UM1AI068636), the President’s Emergency Plan for AIDS, and the AIDS Research Institute at the University of California San Francisco. The content is solely the responsibility of the authors.

\vspace{2em}

\textbf{Conflicts of Interest:} The authors report no conflict of interest in this work. 

\vspace{2em}

\textbf{Acknowledgments:} We thank the Ministries of Health of Uganda and Kenya; our research and administrative teams in San Francisco, Uganda, and Kenya; our collaborators and advisory boards; and, especially, all the communities and participants involved. We also thank Dr. Diane Havlir and Dr. Maya L. Petersen, who together with Dr. Moses R. Kamya are the MPIs of the SEARCH collaboration. 

\vspace{2em}

\textbf{Data and Code Availability:} 
A de-identified dataset and computing code sufficient to reproduce the study findings will be made available following approval of a concept sheet summarizing the analyses to be done. Further inquiries can be directed to the SEARCH Scientific Committee at douglas.black@ucsf.edu.

\newpage

\begin{center}
{\LARGE\textbf{Causal Inference with Missing Exposures and Missing Outcomes}} \\[1.5cm]
\end{center}

\begin{abstract} 

Missing data are ubiquitous in public health research. When estimating causal effects, there are well-established methods to address bias to due missing outcomes. Commonly, causal estimands are defined under hypothetical interventions to ``set'' the exposure \emph{and} to prevent missingness. We demonstrate how this framework can be extended to missing exposures. We further extend this framework to incorporate missingness on the baseline outcome, which induces missingness on the population of interest (e.g., persons at-risk). To do so, we highlight Counterfactual Strata Effects,  a general class of causal estimands where the focus population is subject to missingness and/or impacted by the exposure. They are termed such because the estimand involves conditioning on a counterfactual variable.
For each setting, we present the causal model, relevant counterfactuals, causal estimand, and identification result. We demonstrate with a real-data example to investigate the effect of alcohol consumption on the risk of incident tuberculosis (TB) infection in rural Uganda.   We highlight the use of TMLE with Super Learner for estimation and inference and discuss the practical consequences of our approach.

\end{abstract}


\newpage  


\section*{Introduction}

Missing data affect the integrity of analyses across the spectrum of public health research, including surveillance studies to estimate disease prevalence and randomized trials to establish efficacy of new medical products.\citep{little2012prevention,wells2013strategies,sterne2009multiple,moreno2018canonical,balzer2020far,cole2023missing,juul2024missing,medcalf2024addressing}  There is rich history of methods research to address the potential for bias when participants with measured outcomes differ meaningfully from those with missing or censored outcomes.\citep{robins1986new,PearlCausality,van2011targeted,whatif, horvitzGeneralizationSamplingReplacement1952,rubin1976inference, MarkRobins2003,Robins2000,Taubman2009,schnitzer2014effect,van2018targeted, young2020causal} The Causal Roadmap provides one such approach.\cite{Petersen2014roadmap,dang2023_roadmap,Dang2023,gruber_2024}
The standard implementation of the Roadmap with censored outcomes is as follows. First, we specify our causal question: how would expected outcomes differ if all did versus did not have the exposure of interest \emph{and} censoring were prevented. Then, we specify a causal model, such as a directed acyclic graph (DAG) or non-parametric structural equation model (NPSEM), to represent the data generating process for the baseline and time-varying confounders, exposures, and outcomes.\cite{PearlCausality,van2011targeted}  Third, we intervene on the causal model to generate counterfactual outcomes under hypothetical interventions to set the exposures and ensure outcome measurement.  Fourth, we evaluate whether the corresponding causal estimand can be identified (i.e., if sequential exchangeability and positivity hold).\citep{robins1986new,PearlCausality,van2011targeted,whatif}
Fifth, we specify the statistical estimand, which is often a complex function of the observed data distribution (i.e., not equal to single regression coefficient).
Sixth, we conduct estimation and inference with inverse weighting, standardization, or doubly robust methods, such as targeted minimum loss-based estimation (TMLE).\cite{van2011targeted}
Finally, we conduct sensitivity analyses and appropriately interpret the results.

\vspace{2em}
Here, we discuss how this framework can be extended to address multi-source missingness in real-world (i.e., observational) settings. To do so, we present a series of causal models and identification results, building from simple to complex. First, we  discuss how the Roadmap approach for missing and censored outcomes can be extended to address missingness on  the exposure. Second, we present how this approach can be applied with missing exposures and outcomes. Finally, we address missingness on the exposure, the baseline outcome, and endline outcome.
Concretely, suppose we are interested in studying the incidence of some disease. Our focus population would be persons who are at-risk of the developing the outcome and are, thereby, disease-free at baseline. In this setting, our incidence estimates would be subject to bias if our analyses did not account for differential missingness of outcomes at baseline.  We provide a framework for explicitly defining, identifying, and estimating parameters in such  scenarios, overall and when the exposure influences the baseline risk of the outcome. To do so, we use Counterfactual Strata Effects: a general class of causal estimands defined for a group which is subject to missingness and/or influenced by the exposure.\citep{balzer2017evaluation,petersen_association_2017, Balzer2018SAP,havlir_hiv_2019,balzer2020far, Balzer2021twostage, nugent2023blurring,  marquez2024community, petersen_eurosim_2024, gupta_when_2024,nakato2025measurementmediateseffect}

\vspace{2em}

We illustrate practical relevance with SEARCH-TB, which aimed to evaluate the effect of alcohol use on incident tuberculosis (TB) infection.\cite{abbott2024incident} SEARCH-TB was a sub-study of SEARCH, a cluster randomized trial to evaluate a community-based approach to  Universal HIV Test-and-Treat in rural Kenya and Uganda (2013-2017; NCT01864603).\citep{havlir_hiv_2019} At baseline in SEARCH, we conducted a rapid census and then population-based measurement of sociodemographic factors (e.g.,  age, sex, mobility, and alcohol use) and HIV.\cite{chamie_hybrid_2016} Due to high costs and complex logistics, evaluation of incident TB infection was limited to eastern Ugandan and intentionally enriched for persons with HIV.\citep{Marquez2020, marquez2024community} Specifically, we over-sampled households with at least one adult with HIV. Then we aimed to administer tuberculin skin tests to residents of the sampled households. One year later, we aimed to administer follow-up tests to participants who  tested negative at baseline in the sub-study.  Here, we demonstrate  the methods used to address several real-world challenges arising in SEARCH-TB: confounding, missingness on the exposure of interest (self-report of any alcohol use), missingness on the baseline outcome (defining who was at-risk of TB), and missingness on the final outcome (defining who acquired TB).

\section*{Building in Complexity}

Many studies feature only a subset of the challenges described above. We provide causal models and identification results for a series of hypothetical studies with increasing complexity. For simplicity, we focus on defining and identifying causal estimands under a single level of the exposure, but our results naturally generalize to causal effects defined in terms of contrasts of counterfactual outcome distributions under two levels of the exposure (i.e., the average treatment effect or causal risk ratio). In Appendix S1, we review  the classic ``point-treatment'' problem, where we have measured confounding by baseline covariates $L$, a binary exposure $A$ occurring at single time-point, and an outcome $Y$ occurring at the study's close. Appendix S2 provides a brief review of approaches to handle missingness on the baseline confounders $L$.

\subsection*{Missing Outcomes}

\begin{figure}[H]
    \centering
    \includegraphics[width=0.75\textwidth]{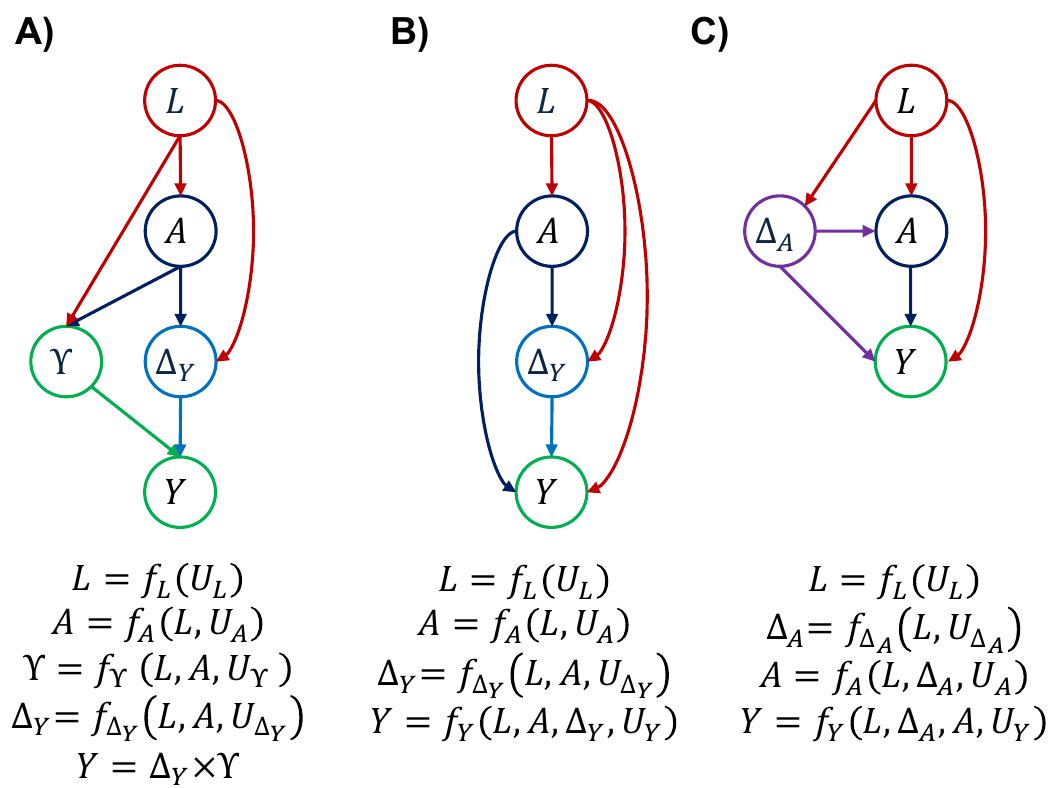}
    \caption{Causal models with missingness on the outcome (Panels A-B)  or missingness on the exposure (Panel C): $L$=baseline covariates, $\Delta_A$=indicator of exposure measurement, $A$=observed exposure, $\Upsilon$=underlying outcome, $\Delta_Y$=indicator of outcome measurement, and $Y$= observed outcome. For ease of presentation, the models are shown without dependence between the unmeasured variables, which are omitted  on the graph.}
    \label{fig:fig_panel}
\end{figure}

Consider a study of the effect of alcohol use on incident TB infection. Suppose that we have a representative cohort of persons without TB at baseline, but some participants did not test for TB at the end of follow-up. Let $\Delta_Y$ be an indicator of outcome measurement. If $\Delta_Y=1$ for a participant, we observe their outcome $Y$ as usual. However, if $\Delta_Y=0$, their outcome $Y$ is not observed. Figure~\ref{fig:fig_panel} provides two possible causal models to represent such a study. In Panel A, we use $\Upsilon$ to represent the underlying value of the outcome, and define the observed outcome as the product of its measurement indicator  and underlying value: $Y=\Delta_Y \times \Upsilon$. Thus, $Y$ equals its underlying value $\Upsilon$ when it's measured (i.e., when $\Delta_Y=1$) and is zero otherwise (i.e., when $\Delta_Y=0$). In Panel B, we omit the underlying outcome and directly represent the causal model in terms of the observed data: $O=(L,A,\Delta_Y, Y)$.  In the latter presentation, the measurement indicator $\Delta_Y$ again influences the value of the observed outcome $Y$, because the outcome is missing if $\Delta_Y=0$. Both approaches lead us  to same identification assumptions and statistical estimands. Therefore, to minimize notation and mirror the long-standing literature on censoring, we use the latter representation in the remainder of the article.\citep{robins1994estimation, Robins2000,Bang&Robins05,Taubman2009,van2011targeted,schnitzer2014effect,van2018targeted, young2020causal,whatif}

\vspace{2em}

To define causal effects when the outcome is subject to missingness, we consider counterfactuals indexed by both the exposure and the outcome measurement indicator. Specifically, let $Y^*=Y^{A=a, \Delta_Y=1}$ be the counterfactual outcome for a given participant if, possibly contrary-to-fact, their exposure were at level $A=a$ and missingness on the outcome prevented. Then our causal estimand  $\mathbb{E}(Y^*)$ is the expected counterfactual outcome if all participants had exposure $A=a$ and their outcomes measured. To identify this causal estimand and express it as a function of the observed data distribution, we would need the baseline covariates $L$ to be sufficient to control for confounding and differential outcome missingness. 
Specifically, we need $L$ to capture all the common causes of the exposure and outcome, and, among those with the exposure of interest, all the common causes of the outcome and its measurement. These assumptions can be represented as $Y^* \perp A \mid L$ and $Y^* \perp \Delta_Y \mid A=a, L$, respectively. These conditions are  often referred to as the ``sequential randomization assumption'' or ``sequential exchangeability'' and can be evaluated graphically through the sequential backdoor criterion.\citep{robins1986new,PearlCausality,van2011targeted,whatif} We also need a positive probability of being exposed to level $A=a$ within all possible values of the confounders and, among those with the exposure of interest, a positive probability of outcome measurement within all possible values of the confounders: $\mathbb{P}(A=a \mid L) > 0$ a.e. and $\mathbb{P}(\Delta_Y \mid A=a, L)>0$ a.e., respectively. If these assumptions hold, we have equivalence between our causal estimand $\mathbb{E}(Y^*)$ and the statistical estimand given by the G-computation formula: $\mathbb{E}[\mathbb{E}(Y\mid \Delta_Y=1, A=a, L)]$.\cite{robins1986new}

\subsection*{Missing Exposures}

We now demonstrate how the above approach for missing outcomes can be applied to missing exposures. Continuing our running example, suppose that among our representative cohort of persons without TB at baseline, some participants did not answer questions about their alcohol use. Let $\Delta_A$ be an indicator that a participant has their exposure  measured. If $\Delta_A=1$ for a participant,  we observe their exposure $A$ as usual. However, if $\Delta_A=0$ for a participant, their exposure $A$ is missing. The DAG and NPSEM for such a study are given in Figure~\ref{fig:fig_panel}C. To define causal effects when the exposure is subject to missingness, we consider counterfactuals indexed by both the exposure and its measurement indicator. Now, let $Y^*=Y^{\Delta_A=1,A=a}$ be the counterfactual outcome for a given participant if, possibly contrary-to-fact, their exposure were measured and at level $A=a$. 

\vspace{2em}

To identify the expected counterfactual outcome $\mathbb{E}(Y^*)$ and  express it as a function of the distribution of the observed data $O=(L, \Delta_A, A,Y)$, we need analogous conditions as the prior subsection. Specifically, we need $L$ to be sufficient to control for differential exposure missingness and for confounding (among those with measured exposures): $Y^* \perp \Delta_A \mid L$ and  $Y^* \perp A \mid \Delta_A=1, L$, respectively. 
We also need the two analogous assumptions on data support:  $\mathbb{P}(\Delta_A=1 \mid L) > 0$ a.e. and $\mathbb{P}(A=a \mid \Delta_A=1, L) > 0$  a.e. If these four assumptions hold, we can rewrite $\mathbb{E}(Y^*)$ as the statistical estimand $\mathbb{E}[\mathbb{E}(Y\mid A=a, \Delta_A=1, L)]$ with proof in Appendix S3.1. 

\subsection*{Missing Exposures and Outcomes}

\begin{figure}[H]
    \centering
    \includegraphics[width=0.5\textwidth]{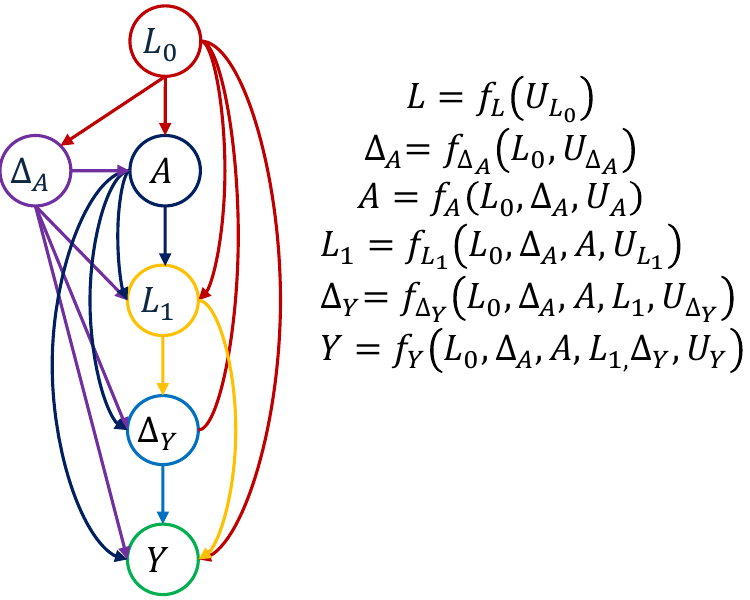}
    \caption{Causal graph and corresponding non-parametric structural equation model  with missingness on the exposure and the outcome: $L_0$=baseline covariates, $\Delta_A$=indicator of exposure measurement, $A$=observed exposure, $L_1$=time-varying covariates, $\Delta_Y$=indicator of outcome measurement, and $Y$=observed outcome. For ease of presentation, the models are shown without dependence between the unmeasured variables, which are omitted  on the graph.}
    \label{fig:fig_miss_a_y}
\end{figure}

We now combine our two challenges: missing exposures and missing outcomes. Continuing our running example, suppose that among our cohort of persons at-risk of TB, some participants did not answer questions about their alcohol use and some participants did not test for TB at the end of follow-up.
To reflect this data generating process, we introduce new notation to reflect the longitudinal setting. Let $L_0$ be baseline covariates and $L_1$ be additional covariates collected after the exposure but before outcome ascertainment. Figure~\ref{fig:fig_miss_a_y} provides the causal models for such a study. 

To define the causal effect when the exposure and outcome are subject to missingness, we now consider counterfactuals indexed by the exposure and two measurement indicators. Specifically, let $Y^*=Y^{\Delta_A=1, A=a, \Delta_Y=1}$ denote the counterfactual outcome under hypothetical interventions to ensure exposure measurement, ``set'' the exposure level to $A=a$, and ensure outcome measurement. To identify the expected counterfactual outcome $\mathbb{E}(Y^*)$ and express it as a function of the distribution of the observed data $O=(L_0, \Delta_A, A, L_1, \Delta_Y, Y)$, we now need to account for the post-baseline covariates $L_1$, which act as time-dependent confounders.  Specifically, $L_1$ are mediators of the exposure-outcome relationship, while ``confounding'' the measurement-outcome relationship. Therefore, we rely on sequential randomization/exchangeability and find a set of covariates that satisfies the backdoor criterion for each ``intervention'' node given the observed past.\cite{robins1986new}  As before, we need that the baseline covariates $L_0$ are sufficient to control for missing exposures and  for confounding. In other words, we need the analogous identification assumptions given in the prior subsection. Additionally, we need that among participants with measured exposures at the level of interest (i.e., $\Delta_A=1$ and $A=a$),
the baseline and time-varying covariates ($L_0,L_1)$ capture all the common causes of outcomes and their measurement as well as a positive probability of outcome measurement within all possible values of the baseline and time-varying covariates: $Y^* \perp \Delta_Y \mid L_1, A=a, \Delta_A=1, L_0$ and $\mathbb{P}(\Delta_Y=1 \mid L_1, A=a, \Delta_A=1, L_0) > 0$ a.e., respectively. If these assumptions hold, we can rewrite $\mathbb{E}(Y^*)$ in terms of the longitudinal G-computation formula: $\mathbb{E}\{\mathbb{E}[\mathbb{E}(Y\mid \Delta_Y=1, L_1, A=a, \Delta_A=1, L_0) \mid A=a, \Delta_A=1, L_0)]\}$, shown in terms of iterated expectations and with proof in Appendix S3.2.\cite{robins1986new, Bang&Robins05, vanderLaan2012towertmle} %

\subsection*{Counterfactual Strata Effects}

\begin{figure}[H]
    \centering
    \includegraphics[width=0.75\textwidth]{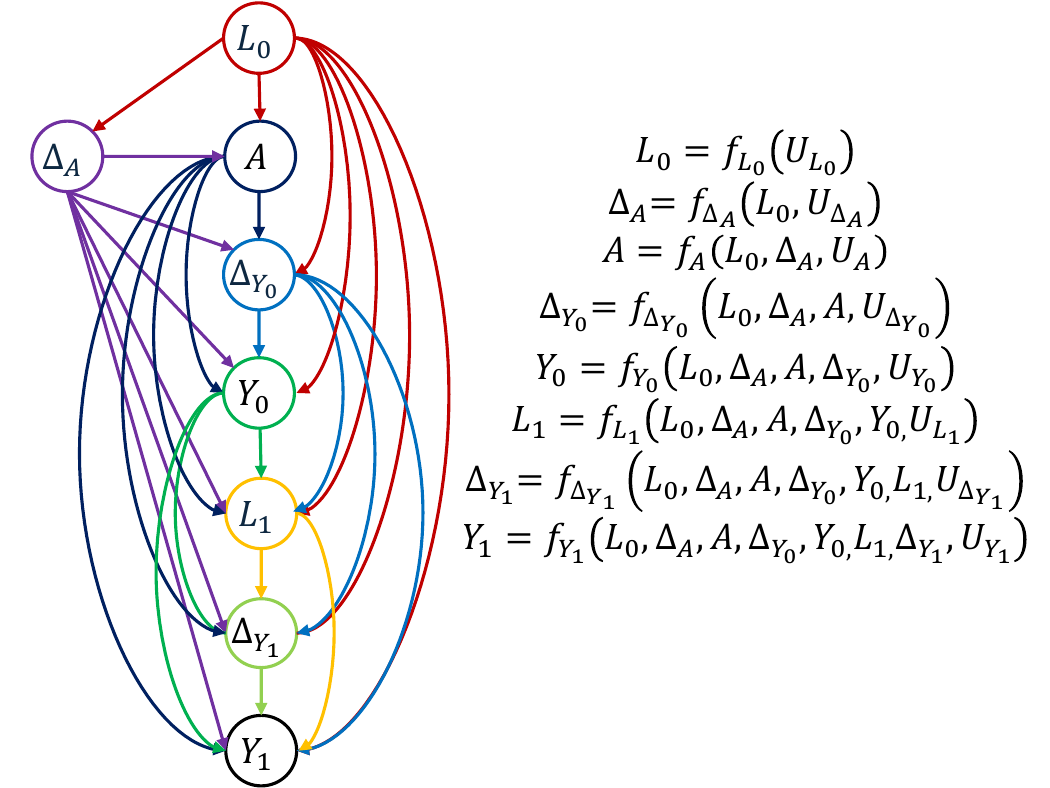}
    \caption{Causal graph and corresponding non-parametric structural equation model  with missingness on the exposure and the outcome at baseline and at endline: $L_0$=pre-exposure  covariates, $\Delta_A$=indicator of exposure measurement, A=exposure, $\Delta_{Y_0}$=indicator of outcome measurement at baseline, $Y_0$=baseline outcome, $L_1$=time-dependent covariates, $\Delta_{Y_1}$=indicator of outcome measurement at endline, and $Y_1$= outcome at endline. For ease of presentation, the models are shown without dependence between the unmeasured variables, which are omitted  on the graph. The time-ordering reflects SEARCH-TB, the real-data example; for an alternative, see Appendix S4.
    }
    \label{fig:overall_dag}
\end{figure}

We now introduce missingness on the outcome at baseline, which induces missingness on the focus population. 
Continuing our running example, suppose that we did not reach all for initial TB testing.  Then our longitudinal cohort of participants may not be representative of all with TB-risk (i.e., subject to selection bias). 
To reflect this challenge, we update our notation to have multiple outcome measures. Let $\Delta_{Y_0}$ be an indicator of outcome measurement at start of follow-up (hereafter ``baseline'') and $\Delta_{Y_1}$ be an indicator of outcome measurement at the end of follow-up (hereafter ``endline''). Let $Y_0$ and  $Y_1$ denote the corresponding outcomes. 
The corresponding causal models can be found in Figure~\ref{fig:overall_dag}. The time-ordering reflects the study protocol and procedures in SEARCH-TB, our real-data example. Specifically, the pre-exposure covariates $L_0$ and alcohol use $A$ were measured in the parent study and before baseline TB status $Y_0$ in the sub-study.\cite{Marquez2020, marquez2024community,abbott2024incident} As a result, the exposure $A$ can impact who has prevalent TB at baseline ($Y_0=1$) and, thereby, who is at-risk of TB at baseline ($Y_0=0$). 
As demonstrated in Appendix S4, an analogous approach can be taken when  baseline outcome measurement occurs before exposure measurement. In both settings, we need to address the potential for bias due to differential measurement of the baseline outcome, which defines our focus population. In other words, the causal estimands of interest are Counterfactual Strata Effects.

To define the causal estimand in the setting shown in Figure~\ref{fig:overall_dag}, we first consider the counterfactual outcome \emph{at baseline} under hypothetical interventions to ensure exposure measurement, ``set'' the exposure level to $A=a$, and ensure outcome measurement at baseline: $Y_0^* = Y_0^{\Delta_A=1, A=a, \Delta_{Y_0}=1}$. Additionally, we consider the counterfactual outcome \emph{at endline} under the prior interventions as well as a hypothetical intervention to ensure outcome measurement at endline among those known to be at-risk at baseline. Practically, only participants who tested TB-negative at baseline ($\Delta_{Y_0}=1$ and $Y_0=0$) would be approached for re-testing at endline. Thereby, this final intervention is a dynamic or personalized.\citep{balzer2017evaluation, hernan2006comparison, van2007causal,robins2008estimation} We  ``set'' $\Delta_{Y_1}$ equal to one if $Y_0=0$ and to zero otherwise. For simplicity, denote the resulting counterfactual outcome as
$Y_1^* = Y_1^{\Delta_A=1, A=a,\Delta_{Y_0}=1, \Delta_{Y_1}=1}$. 

\vspace{2em}

Now we can precisely define the causal estimand in terms of the following conditional probability, which captures the counterfactual incidence of the outcome among all at-risk at baseline:  $$\mathbb{P}(Y_1^*=1 \mid Y_0^*=0)$$ Due to conditioning on a counterfactual variable, such parameters are sometimes called ``Counterfactual Strata Effects'',  which are defined by Nakato et al. as  ``causal estimands where the outcome is only relevant for a group whose membership is subject to missingness and/or impacted by the exposure''.\citep{nakato2025measurementmediateseffect}
They are a general class of causal estimands of the form:
$$\mathbb{E}(Y_1^* \mid Y_0^*=y_0^*)$$
where the definition of the counterfactuals ($Y_0^*, Y_1^*$) varies by the application.
Over the past decade,
these estimands have been specified to evaluate intervention effects on the cumulative incidence of outcomes and care/prevention cascades in cluster randomized trials.\citep{balzer2017evaluation,petersen_association_2017, Balzer2018SAP,havlir_hiv_2019,balzer2020far, Balzer2021twostage, nugent2023blurring,  marquez2024community, petersen_eurosim_2024,gupta_when_2024,nakato2025measurementmediateseffect}
To our knowledge, this is their first use in observational studies and with additional missingness on the exposure. (Appendix S5 provides a discussion of how Counterfactual Strata Effects relate to Principal Strata Effects.)

\vspace{2em}

To facilitate identification, we use probability rules to re-express  the conditional probability as 
\begin{align}
\mathbb{P}(Y_1^*=1 \mid Y_0^*=0) = \frac{\mathbb{P}(Y_1^*=1, Y_0^*=0)}{\mathbb{P}(Y_0^*=0)}
\label{eq:causal_parameter}
\end{align}
Then given  the  observed data $O=(L_0, \Delta_A, A, \Delta_{Y_0}, Y_0, L_1, \Delta_{Y_1}, Y_1)$, we can identify the denominator and the numerator.
The denominator $\mathbb{P}(Y_0^*=0)$ represents the counterfactual prevalence of not having TB at baseline 
--- under exposure $A=a$ and under complete measurement of the exposure and baseline outcome ($\Delta_A=\Delta_{Y_0}=1)$.
The causal structure for this parameter is analogous to that of Section 3.2, but with an additional measurement indicator for the baseline outcome. Therefore, under analogous assumptions, we can identify $\mathbb{E}(Y_0^*)=\mathbb{E} \big[ \mathbb{E}(Y_0 \mid \Delta_{Y_0}=1, A=a, \Delta_{A}=1, L_0) \big]$ with proof in Appendix S3.3. Since we are interested the counterfactual probability of being at-risk at baseline $\mathbb{P}(Y_0^*=0)$, our statistical estimand for the denominator becomes $1- \mathbb{E} \big[ \mathbb{E}(Y_0 \mid \Delta_{Y_0}=1, A=a, \Delta_{A}=1, L_0) \big]$. Altogether, this statistical estimand accounts for differential measurement of the exposure and baseline TB status as well as the impact of exposure on baseline TB risk. We again refer to Appendix S4 for an alternative time-ordering where the exposure does not impact the outcome at baseline.

\vspace{2em}


In our final causal parameter (Eq.~\ref{eq:causal_parameter}), the numerator $\mathbb{P}(Y_1^*=1, Y_0^*=0)$ represents the counterfactual probability of acquiring TB --- under exposure $A=a$ and under complete measurement $\Delta_A=\Delta_{Y_0}=\Delta_{Y_1}=1$. For ease of notation, let $Z^*=\mathbb{I}(Y_1^*=1, Y_0^*=0)$ represent the joint indicator of these two counterfactual values. 
To identify $\mathbb{E}(Z^*)=\mathbb{P}(Z^*=1)$, we need analogous assumptions as for the denominator together with the following. Among those known to be at-risk at baseline ($\Delta_{Y_0}=1, Y_0=0$) and with measured exposure of interest ($\Delta_A=1, A=a$):
the baseline and time-varying covariates capture the common causes of the joint  outcome  and endline measurement: 
  $Z^* \perp \Delta_{Y_1} \mid L_1, Y_0=0, \Delta_{Y_0}=1, A=a, \Delta_A=1, L_0$.
We also need there to be a positive probability of endline  measurement within all possible values of $L_0$ and $L_1$: $\mathbb{P}(\Delta_{Y_1}=1 \mid L_1, Y_0=0, \Delta_{Y_0}=1, A=a, \Delta_A=1, L_0) > 0 $ a.e.
Under these assumptions and with proof given in Appendix S3.3, the numerator is identified as
$$ \mathbb{P}(Z^*=1) \ = \ 
\mathbb{E}[  \mathbb{E}\left(   \mathbb{E}(Y_1 \mid \Delta_{Y_1}=1, L_1, Y_0=0, \Delta_{Y_0}=1, A=a, \Delta_A=1, L_0) \mid \Delta_{Y_0}=1, A=a, \Delta_A=1, L_0  \right) ]$$

Putting it all together, our statistical estimand is given by
\begin{eqnarray}
\Psi(\mathbb{P};a) =
    \frac{\mathbb{E}[  \mathbb{E}\left(   \mathbb{E}(Y_1 \mid \Delta_{Y_1}=1, L_1, Y_0=0, \Delta_{Y_0}=1, A=a, \Delta_A=1, L_0) \mid \Delta_{Y_0}=1, A=a, \Delta_A=1, L_0  \right) ]    } { 1-  \mathbb{E} \left[\mathbb{E}(Y_0 \mid \Delta_{Y_0}=1, A=a, \Delta_A=1, L_0)\right]  } 
\label{eq:final}
\end{eqnarray}
for exposure level $A=a$.
Then we can define associations in terms of contrasts $\Psi(\mathbb{P};a)$ at different exposure levels. 
Specifically, we may be interested associations on the difference scale $\Psi(\mathbb{P};1) - \Psi(\mathbb{P};0)$ or on the relative scale
$\Psi(\mathbb{P};1) \div \Psi(\mathbb{P};0)$.

\section*{Statistical Estimation and Inference}

In the previous section, we introduced a series of causal models and identification results of increasing complexity. For the resulting statistical estimands, we could use singly robust estimators, such as standardization (a.k.a., ``G-computation'') or inverse probability weighting (IPW).\citep{robins1986new, horvitzGeneralizationSamplingReplacement1952} Here, we highlight the use of TMLE, which is a doubly robust procedure and asymptotically efficient under certain  conditions.\citep{van2011targeted}
In TMLE, initial estimates of the relevant pieces of  the observed data distribution are  updated to achieve the optimal bias-variance trade-off for the estimand and to solve the efficient influence curve equation.  
Initial estimates 
are often computed via Super Learner, an ensemble machine learning algorithm using  V-fold cross-validation to select an optimal weighted linear combination of predictions  from a library of candidate learners.\citep{van2007super} 
Thereby, TMLE leverages machine learning  to avoid introducing new modeling assumptions during estimation, while supporting valid statistical inference under reasonable conditions.
Notably, for ratio-type estimands corresponding to Counterfactual Strata Effects (Eq.~\ref{eq:final}), we would implement a separate TMLE for the numerator estimand (the joint probability) and the denominator estimand (1- the  baseline prevalence) before combining the results. 

\vspace{2em}

TMLE is an asymptotically linear estimator and is normally distributed in the large data limit.\citep{van2011targeted} 
The estimator minus the estimand behaves like a sample mean in the first order:  $\hat{\Psi} - \Psi = \frac{1}{N}\sum_{i=1}^N D_i + o_P(N^{-1/2})$ where $D_i$ is the influence curve for participant $i=\{1,\ldots,N\}$   and $o_P(N^{-1/2})$ is a second-order remainder term going to zero in probability.\citep{vanderVaart1998} The estimated influence curve is used to calculate standard errors, Wald-type confidence intervals, and p-values. Concretely, a 95\% confidence interval is constructed using $\hat{\Psi} \pm z_{0.0975} \frac{\hat{\sigma}}{\sqrt{n} }$ where $z_{0.975}$ is the critical value at the 97.5th-percentile of the standard normal  and $\hat{\sigma}$ is the standard deviation of the estimated influence curve. For ratio-type estimands (Eq.~\ref{eq:final}), once the influence curves for the numerator and denominator have been estimated, the Delta method provides an estimate of the influence curve for our overall estimand.\cite{vanderVaart1998,Moore2009}
Then to calculate measures of association on the difference, ratio, or odds ratio scale, we apply the Delta method a second time to get inference for these types of functionals (Appendix S6).

\vspace{2em}

To evaluate the practical performance of our approach, we conducted a simulation study reflecting the final causal model (Figure~\ref{fig:overall_dag}). The simulation design and results are detailed in Appendix S7. Briefly, we found that TMLE had negligible bias and achieved nominal confidence interval coverage for the exposure-specific estimands, the risk ratio, and risk difference. All other approaches exhibited meaningful bias and sub-nominal confidence interval coverage.

\section*{Real-Data Example}

The results of SEARCH-TB's investigation of the effect of alcohol use on incident TB infection among adults (15+ years) in rural Eastern Uganda have been previously published.\cite{abbott2024incident} Here, we demonstrate the methods and practical consequences. With our multinational and interdisciplinary team, we worked through the Causal Roadmap.\citep{Petersen2014roadmap,dang2023_roadmap,Dang2023,gruber_2024}
Our causal model reflected the team's knowledge of the study protocol,  study procedures, and epidemiology of TB in the region.
Our adjustment set $L_0$ included the SEARCH trial arm, community indicators, household HIV status, as well as individual-level age, sex, and mobility measures. 
The exposure $A$ was self-report of any alcohol use in response to ``do you drink alcohol?''.
The  covariates and exposure  were measured at the start of the parent SEARCH study (2013-2014), while TB status at   baseline $Y_0$ and endline $Y_1$ was measured subsequently in the sub-study (2015-2017).\cite{havlir_hiv_2019, Marquez2020,marquez2024community}
There were no time-varying covariates $L_1$ measured in SEARCH-TB. 
As detailed in the Statistical Analysis Plan,\cite{abbott2024incident} the causal estimand was the relative risk of incident TB infection: $\mathbb{P}(Y_1^1=1\mid Y_0^1=0)\div(Y_1^0=1\mid Y_0^0=0)$, 
where $Y_0^a$ and $Y_1^a$ denote the counterfactual outcomes at baseline and endline, respectively, under alcohol use $A=a$ and complete measurement. The corresponding statistical estimand was $\Psi(\mathbb{P};1)\div \Psi(\mathbb{P};0)$ with
\begin{eqnarray*}
\Psi(\mathbb{P};a) = \frac{\mathbb{E}[  \mathbb{E}\left(   \mathbb{E}(Y_1 \mid \Delta_{Y_1}=1, Y_0=0, \Delta_{Y_0}=1, A=a, \Delta_A=1, L_0) \mid \Delta_{Y_0}=1, A=a, \Delta_A=1, L_0  \right) ]    } { 1-  \mathbb{E} \left[\mathbb{E}(Y_0 \mid \Delta_{Y_0}=1, A=a, \Delta_A=1, L_0)\right]   } 
\end{eqnarray*}

\vspace{2em}

For the primary analysis, we used TMLE  with Super Learner to combine estimates 
from generalized linear models, 
multivariate adaptive regression splines,  and the mean. 
We conducted influence curve-based inference, accounting for clustering by household (Appendix S8).\cite{laan_estimating_2013,nugent2023blurring} 
In secondary analyses, we considered communities, instead of households, to be the independent unit.  To examine the sensitivity of our results to alternative estimation approaches, we implemented two singly robust approaches for the final statistical estimand (Eq.~\ref{eq:final}):  G-Computation using parametric regressions to estimate the iterated conditional expectations and  IPW using parametric regressions to estimate the weights.\cite{Bang&Robins05}
Finally, to examine the impact of our handling of missing data, we took the following ``na{\"i}ve'' approach: 
(1) subset on participants with complete data $\Delta_A=\Delta_{Y_0}=\Delta_{Y_1}=1$;
(2) further subset on participants known to be at-risk at baseline $Y_0=0$, and
(3) implement TMLE to adjust for confounding and to estimate $\mathbb{E}[\mathbb{E}(Y_1 \mid A=a, L_0 )]$ among the remaining subset of participants.
This approach is inherently flawed, because the corresponding statistical estimand will not equal the wished-for causal estimand ($\mathbb{P}(Y_1^1=1\mid Y_0^1=0) \div (Y_1^0=1\mid Y_0^0=0)$), even if all variables are missing completely at random.
Nonetheless, the approach could be one taken by an analyst aiming to implement a ``complete-case'' analysis while adjusting for confounding.\citep{ross2020complete, mathur2024resurrecting,ghazaleh2021handling}
For statistical inference, G-computation, IPW, and the na{\"i}ve approach accounted for clustering by household.

\vspace{2em}

\begin{figure}[H]
    \centering
    \includegraphics[width=12cm, height=9cm]{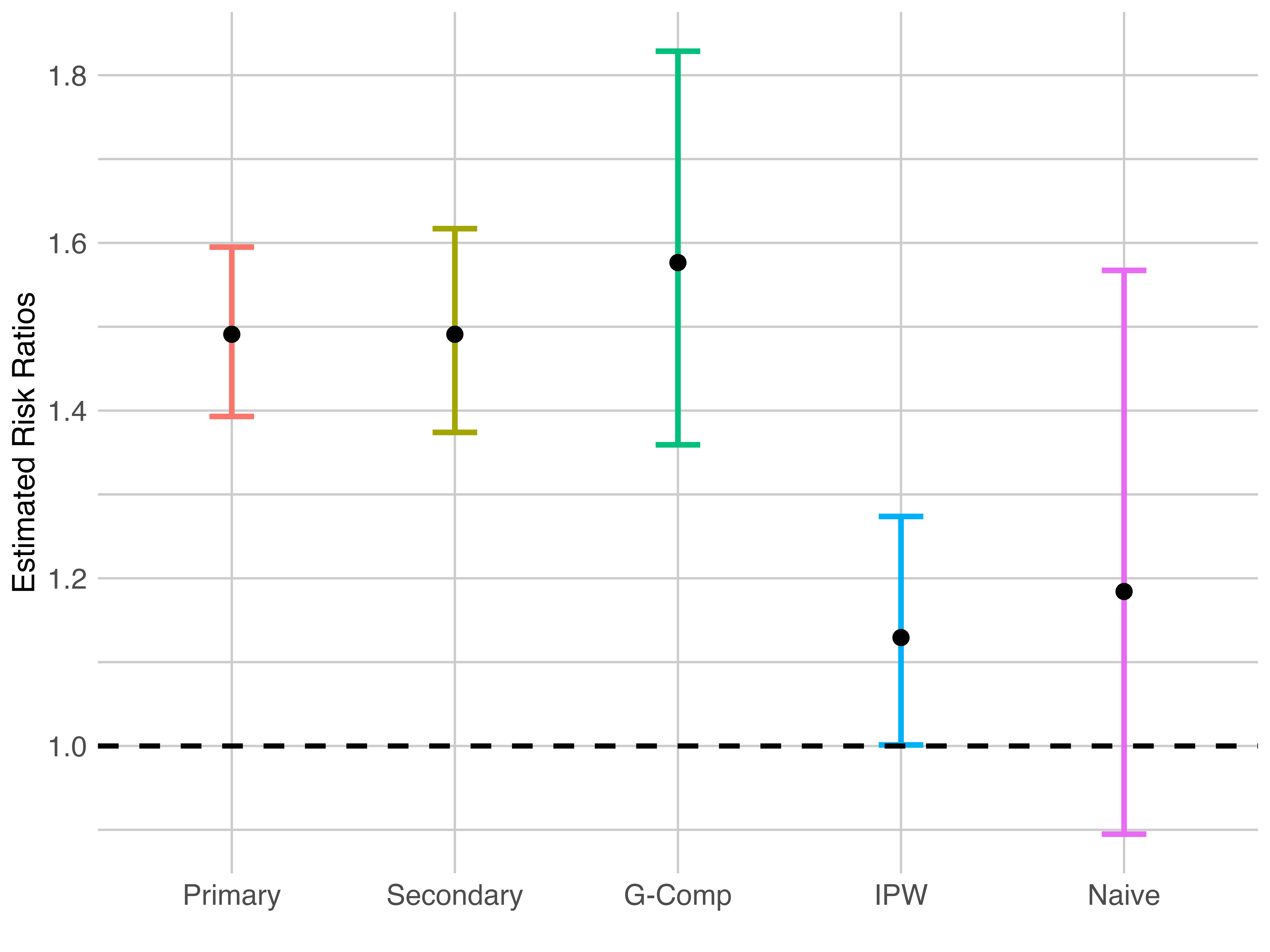}
    \caption{Results from SEARCH-TB for the association of alcohol use on incident tuberculosis (TB) infection: ``Primary'' with TMLE and clustering by household, ``Secondary'' with TMLE and clustering by community, ``G-Comp'' with G-computation, ``IPW'' with inverse probability weighting, and ``Na{\"i}ve'' based on subsetting on those at-risk at baseline and with measured exposures and outcomes at endline.}
    \label{fig:results}
\end{figure}

In the primary analysis using TMLE with clustering by household, we found that  alcohol use was associated with a 49\% increase in the risk of incident TB
: risk ratio (RR)=1.49 (95\%CI: 1.39-1.59).\cite{abbott2024incident} As shown in Figure~\ref{fig:results}, secondary analyses with the community as the independent unit yielded very similar results, despite meaningfully reducing the effective sample size to 9 communities:  RR=1.49 (95\%CI: 1.37-1.62); Appendix S8. In contrast,  G-computation resulted in a larger association with large confidence intervals, while IPW resulted in a smaller association and confidence intervals overlapping the null: RR=1.58 (95\%CI: 1.36-1.83) and RR=1.13 (95\%CI: 1.00-1.27), respectively. 
Finally, after restricting to participants who responded to questions about alcohol use, tested negative at baseline,  and tested again at follow-up, the na{\"i}ve approach was the least precise and resulted in the widest confidence intervals: RR=1.18 (95\%CI: 0.89-1.57).

\section*{Discussion}

We presented causal models, causal estimands, and identification results for a series of prospective studies with increasing levels of missingness.
We reviewed how the Causal Roadmap could be applied to address missing outcomes and presented an extension for missing exposures. We then motivated the use of Counterfactual Strata Effects to precisely define causal estimands when the focus population is subject to missingness and/or impacted by the exposure. Specifically, we considered the common scenario where there is missingness on the outcome at baseline ---  determining who is at-risk of acquiring the outcome by endline. To reflect the real-data example, we allowed for the exposure to influence baseline outcome risk. In Appendix S4, we presented an alternative time-ordering with baseline outcomes were measured prior to the exposure. Both scenarios resulted in Counterfactual Strata Effects, which we then identified after re-expressing the conditional estimand in terms of a ratio.
In addition to incidence studies,\cite{marquez2024community,nugent2023blurring,gupta_when_2024} Counterfactual Strata Effects commonly arise when studying care cascades: the series of steps from screening, through diagnosis and treatment, to control.\citep{balzer2017evaluation,petersen_association_2017, Balzer2018SAP,havlir_hiv_2019,balzer2020far, Balzer2021twostage, gupta_when_2024, nakato2025measurementmediateseffect}
Indeed, these cascades are a series of conditional probabilities --- each subject to missingness and each potentially impacted by the exposure or intervention strategy.
Altogether, the Causal Roadmap approach facilitates formal definition and identification of the corresponding causal estimand in a variety of settings. For estimation and inference, we highlighted the use of TMLE with Super Learner  to robustly and efficiently estimate the corresponding statistical estimands.

\vspace{2em}

There are several advantages to our framework for handling multi-source missingness, which may arise by design or despite best intentions of study investigators. 
First, we demonstrated the use of causal models to reflect the data generating process, generate counterfactuals, and aid in the identification of the corresponding causal estimand.
Furthermore, unlike the m-DAG approach which aids in identifying the entire joint distribution of the observed data,\cite{moreno2018canonical} our approach focuses our efforts on the portion of the observed data distribution relevant for the statistical estimand (Appendix S9). 
Our approach also uses data on all participants, improving efficiency relative to approaches excluding participants with missing data on the relevant variables (Figure~\ref{fig:results}; Appendix S.7). Finally, our approach leads us to statistical estimands that can be robustly and rigorously estimated with modern methods, such as TMLE. 

\vspace{2em}

There are several limitations to our work. First, we focused on cross-sectional or prospective studies. Thus, we did not cover scenarios where the outcome impacts the measurement of other variables. Such scenarios would arise in case-control studies and have been addressed in prior literature.\citep{RothmanModern,rose2011targeted, zhang2016causal,kennedy2020efficient,ghazaleh2021handling, qiu2026efficient}
Second,  we did not provide an exhaustive set of causal models and identification results for all possible studies; however, our approach is generalizable and covers many scenarios  arising in public health. Third, 
we did not consider multiple imputation, which is a common approach for missing data.\cite{Rubin1987}
However, foundational work is needed to investigate the assumptions, implementation, and performance of multiple imputation in settings with (1) missingness on the exposure, (2) missingness on the baseline outcome,
(3) missingness in the final outcome, (4) confounding,
and (5) dependence among study participants. Indeed, Tompsett et al. suggested that multiple imputation cannot be used in settings with missingness on the focus population (termed ``missing eligibility data'') and missingness on the exposure.\cite{tompsett2023target}
Finally, we relied on various versions of the sequential exchangeability assumption for both confounding and missingness.
In practice, data may be missing as a result of unobserved variables, and  
we may need to collect additional data as well as  conduct sensitivity/bias  analyses.\citep{lash_bias,cornelisz2020addressing,Gruber2024} Nonetheless, 
even when a ``causal gap'' remains, we have a framework to define  a statistical estimand, which is aligned with our research question, and appropriately addresses many real-world complications.\citep{Petersen2014roadmap, dang2023_roadmap, Dang2023, gruber_2024}

\newpage

\section*{Supplementary Materials for ``Causal Inference with Missing Exposures and Missing Outcomes''}

Contents:
\begin{itemize}
\item Appendix S1: The classic point-treatment problem 
\item Appendix S2: Missing confounders
\item Appendix S3: Proofs
\item Appendix S4: Alternative time-ordering
\item Appendix S5: Principal Strata Effects
\item Appendix S6: More on the Delta method
\item Appendix S7: Finite sample simulation study
\item Appendix S8: Accounting for outcome dependence
\item Appendix S9: Missingness DAGs
\end{itemize}


\section*{Appendix S1: The classic point-treatment problem}

\begin{figure}[H]
    \centering
  \includegraphics[width=0.75\textwidth]{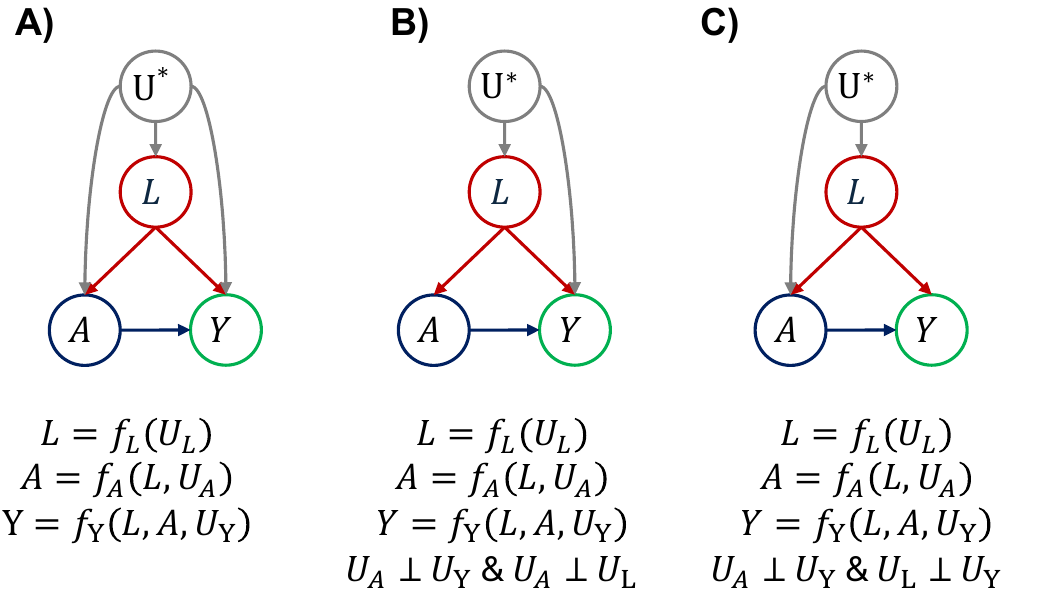}
    \caption{Causal models for a classic point-treatment problem with complete measurement of the baseline covariates $L$, the exposure $A$, and the outcome $Y$. On the directed acyclic graph, $U^*$ represents unmeasured common causes of at least two variables in $(L,A,Y)$. In panel A provides the causal models under no assumptions about the unmeasured factors. Panels B and C are compatible with the no unmeasured confounders assumption.}
    \label{fig:fig_no_miss}
\end{figure}

We consider the classic ``point-treatment'' problem, where we have measured confounding by baseline covariates $L$, a binary exposure $A$ occurring at single time-point, and an outcome $Y$ occurring at the study's close. This could represent a study of the effect of alcohol use ($A$) on incident TB infection ($Y$) among a representative cohort of persons without TB at baseline. The directed acyclic graph (DAG) and non-parametric structural equation model (NPSEM) for such a study are given in Figure~\ref{fig:fig_no_miss}A. 

\vspace{2em}

Under interventions on the causal model, we  generate counterfactual outcomes corresponding to the research question of interest. Specifically, let $Y^{a}$ be the counterfactual outcome for a given participant if, possibly contrary-to-fact, they had exposure-level $A=a$. Then our causal estimand  $\mathbb{E}(Y^{a})$ is the expected counterfactual  outcome if all  had exposure-level $A=a$. In our running example, $\mathbb{E}(Y^{a})=\mathbb{P}(Y^{a}=1)$ is the counterfactual risk of  incident TB infection with alcohol use $A=a$. 

To identify our causal target parameter and express it as function of the distribution of the observed data $O=(L,A,Y)$, we would need there to be no unmeasured confounding, which corresponds to the assumption that the baseline covariates $L$ capture all the joint causes of the exposure $A$ and outcome $Y$. This condition is called ``the randomization assumption'' or ``exchangeability'', can be evaluated with the backdoor criterion, and can be represented as $Y^a \perp A \mid L$.\citep{robins1986new,PearlCausality,van2011targeted,whatif}
Concretely, $\mathbb{E}(Y^{a})$ is not identified in Figure~\ref{fig:fig_no_miss}A because there is an unmeasured common cause, represented by $U^*$, of the exposure $A$ and outcome $Y$. In Figure~\ref{fig:fig_no_miss}B and C, we show two causal models where this assumption  would hold.

\vspace{2em}

Additionally, we need there to be a non-zero probability of having the exposure in all possible values of $L$:  $\mathbb{P}(A=a \mid L) > 0$ a.e.  This is a condition on data support and known as the ``positivity assumption''. 
We note that the consistency assumption, stating that the counterfactual outcome $Y^a$ equals the observed outcome $Y$ under exposure $A=a$, holds by design when we define counterfactuals through interventions on the causal model.
Under the exchangeability and positivity assumptions, our causal target is equal to the statistical estimand given by the G-computation formula: $\mathbb{E}[\mathbb{E}(Y\mid A=a, L)]$.\cite{robins1986new}  Even if these assumptions are not reasonable (e.g., Figure~\ref{fig:fig_no_miss}A reflects reality), we still have well-defined statistical estimand, on which we can focus our estimation efforts.
In other words, we  still proceed with estimation and inference  $\mathbb{E}[\mathbb{E}(Y\mid A=a, L)]$ , while appropriately accounting for lack of identification in our interpretation.\cite{Petersen2014roadmap,dang2023_roadmap,gruber_2024,Wong2021,Dang2023}
For ease of presentation in main text and following Appendices, we provide the causal models without dependence between the unmeasured factors. 

\newpage

\section*{Appendix S2: Missing Confounders}

Throughout, we also assume that the covariates are completely measured. While missingness on covariates is common, a detailed discussion is beyond the scope of this manuscript, which focuses on defining, identifying, and estimating causal effects in settings with missing exposures and missing outcomes. Here, we provide a brief review of methods to handle missing confounders. Common approaches include dropping covariates subject to missingness, conducting a complete (available) case analysis, including a missingness indicator, and using multiple imputation.\cite{Groenwold_missing_2012,xu2021missing,williamson2024assessing, qiu2026efficient}  Excluding covariates subject to missingness will induce bias if those covariates are confounders. A complete case analysis, where observations with missing data are excluded, is guaranteed to reduce power and will be biased unless missingness is completely at random. In the missing indicator method,  a dummy variable is included to indicate the covariate is measured/missing. Recently, the missingness indicator method was shown to perform similarly to multiple imputation (MI) in many settings.\cite{xu2021missing} MI uses observed variables to predict the values of missing ones and has traditionally relied on strong modeling assumptions, although there is a growing interest in using machine learning.\cite{rubin1976inference, laqueur2022supermice, carpenito2022misl}

\vspace{2em}

Williamson et al. provide an excellent discussion and evaluation of more modern methods for missing confounders, including doubly robust approaches of generalized raking and targeted minimum loss-based estimation (TMLE).\cite{williamson2024assessing} These approaches are more robust to model misspecification and are efficient in certain scenarios. Williamson et al.  also provide guidance on selecting the approach to handle missing confounders based on the bias-variance trade-off.\cite{williamson2024assessing} Recently, Qiu et al. extended this work and showed TMLE, a plug-in estimator, had superior performance to generalized ranking in finite samples.\cite{qiu2026efficient}

\newpage

\section*{Appendix S3: Proofs}

In the following, we provide proofs for the identification results. To match the applied example, we focus on binary outcomes, but our results generalize to all outcome-types. For simplicity we focus on categorical covariates, but our summations generalize to integrals for continuous covariates. 

\subsubsection*{Appendix S3.1: Missing exposures (Figure 1C in the main text)}

Let $Y^*=Y^{\Delta_A=1, A=a}$. Then we have equivalence between our wished-for causal estimand and the corresponding statistical estimand under the following identifability assumptions:

\begin{eqnarray*}
\mathbb{P}(Y^* = 1)&=&\sum_{l}\mathbb{P}(Y^*=1 \mid L=l)\mathbb{P}(L=l) \\
&& \text{by } Y^* \perp \Delta_A \mid L \\ 
&=& \sum_{l}\mathbb{P}(Y^*=1 \mid \Delta_A=1, L=l)\mathbb{P}(L=l) \\ 
&& \text{by } Y^* \perp A \mid \Delta_A=1, L \\ 
&=& \sum_{l}\mathbb{P}(Y^*=1 \mid A=a, \Delta_A=1, L=l)\mathbb{P}(L=l) \\
&& \text{by the consistency assumption} \\ 
&=& \sum_{l}\mathbb{P}(Y=1 \mid A=a, \Delta_A=1, L=l)\mathbb{P}(L=l) \\
&=& \mathbb{E}\big[ \mathbb{E}(Y\mid A=a, \Delta_A=1, L)\big]
\end{eqnarray*}

We again note the consistency assumption holds by our definition of counterfactual outcomes as being derived through interventions on the causal model.
For the corresponding statistical estimand to be well-defined, we also need the following positivity assumptions: $\mathbb{P}(\Delta_A=1|L)>0$  a.e.  and $\mathbb{P}(A \mid \Delta_A=1, L)>0$ a.e.. We note that the above approach can be simplified. Since there is no intervening nodes (variables) between $\Delta_A$ and $A$, we can instead consider a joint intervention variable $\alpha=\mathbb{I}(\Delta_A=1,A=a)$. Under analogous assumptions $Y^* \perp \alpha \mid L$ and $\mathbb{P}(\alpha=1|L)>0$, we have the same identification result.

\subsubsection*{Appendix S3.2: Missing Exposures and Outcomes (Figure 2 in the main text)}

Let $Y^*=Y^{\Delta_A=1, A=a, \Delta_Y=1}$. Then we have equivalence between our wished-for causal estimand and the corresponding statistical estimand under the following identifability assumptions:
\begin{eqnarray*}
\mathbb{P}(Y^{*} = 1)&=&\sum_{l_0}\mathbb{P}(Y^*=1 \mid L_0=l_0)\mathbb{P}(L_0=l_0) \\
&& \text{by } Y^* \perp \Delta_A \mid L_0 \text { and } Y^* \perp A \mid \Delta_A=1, L_0 \\ 
&=& \sum_{l_0}\mathbb{P}(Y^*=1 \mid  A=a,\Delta_A=1,L_0=l_0)\mathbb{P}(L_0=l_0) \\
&& \text{by } Y^* \perp \Delta_Y \mid L_1, A=a, \Delta_A=1, L_0 \\ 
&=& \sum_{l_0}\sum_{l_1}\mathbb{P}(Y^*=1 \mid \Delta_Y=1, L_1=l_1, A=a, \Delta_A=1, L_0=l_0) \times \\
&&  \quad \mathbb{P}(L_1=l_1 \mid  A=a, \Delta_A=1, L_0=l_0)\mathbb{P}(L_0=l_0) \\
&=& \mathbb{E}\bigg\{ \mathbb{E}\big[ \mathbb{E}(Y \mid \Delta_Y=1, L_1, A=a, \Delta_A=1, L_0) \mid A=a, \Delta_A=1, L_0) \big] \bigg\}
\end{eqnarray*}
where the inner expectation averages out the outcome $Y$ given the conditioning set, the middle expectation average out the time-varying covariates $L_1$ given the conditioning set, and the outer expectation averages out the baseline covariates $L_0$.
For the corresponding statistical estimand to be well-defined, we also need the following positivity assumptions: $\mathbb{P}(\Delta_Y=1 \mid L_1, A=a, \Delta_A=1, L_0) > 0$ a.e. in addition to the positivity assumptions from the previous section.

\subsubsection*{Appendix S3.3: Missing Exposures and Outcomes at Baseline and Follow-up (Figure 3 in the main text)}

Let $Y_0^*=Y_0^{\Delta_A=1, A=a, \Delta_{Y_0}=1}$ and $Y_1^*=Y_1^{\Delta_A=1, A=a,\Delta_{Y_0}=1, \Delta_{Y_1}=1}$. Recall that we defined the target parameter for this section using Counterfactual Strata Effects:\citep{Balzer2017CascadeMethods,petersen_association_2017,balzer2020far,Balzer2021twostage, nugent2023blurring, petersen_eurosim_2024, gupta_when_2024, havlir_hiv_2019, marquez2024community}
$$\mathbb{P}(Y_1^*=1 \mid Y_0^*=0) \ = \ \frac{\mathbb{P}(Y_1^*=1, Y_0^*=0)}{\mathbb{P}(Y_0^*=0)}$$
Using the form of the target parameter on the right-hand side of the above equation, we proceed by presenting a separate identification result for the numerator and the denominator.

\textbf{Identification proof for the denominator}

Under the following assumptions, which are analogous to Appendix S3.1, we can identify 1 minus the denominator:
\begin{eqnarray*}
\mathbb{P}(Y_0^* = 1)&=&\sum_{l_0}\mathbb{P}(Y_0^*=1 \mid L_0=l_0)\mathbb{P}(L_0=l_0) \\
&& \text{by } Y_0^* \perp \Delta_A \mid L_0 \text { and } Y_0^* \perp A \mid \Delta_A=1, L_0 \text { and } Y_0^* \perp \Delta_{Y_0} \mid A=a, \Delta_a=1, L_0\\ 
&=& \sum_{l_0}\mathbb{P}(Y_0^*=1 \mid \Delta_{Y_0}=1, A=a, \Delta_A=1, L_0=l_0)\mathbb{P}(L_0=l_0) \\
&=& \mathbb{E}\big[\mathbb{E}(Y_0 \mid \Delta_{Y_0}=1, A=a, \Delta_A=1, L_0) \big]
\end{eqnarray*}
along with the corresponding positivity assumptions. We again note that because there are no intervening nodes between $\Delta_A$, $A$, and $\Delta_{Y_0}$, the identification approach can be simplified, as outlined in Appendix S3.1. 

\vspace{1em}

\textbf{Identification proof for the numerator}

Let $Z^*=\mathbb{I}(Y_1^*=1, Y_0^*=0)$. Then under the following assumptions, we can identify the numerator $\mathbb{P}(Y_1^*=1, Y_0^*=0)\ = \ \mathbb{P}(Z^*=1)$.

\begin{eqnarray*}
\mathbb{P}(Z^* = 1)&=&\sum_{l_0}\mathbb{P}(Z^*=1 \mid L_0=l_0)\mathbb{P}(L_0=l_0) \\
&& \text{by } Z^* \perp \Delta_A \mid L_0 \text { and } Z^* \perp A \mid \Delta_A=1, L_0 \text { and } Z^* \perp \Delta_{Y_0} \mid A=a, \Delta_A=1, L_0\\ 
&=& \sum_{l_0}\mathbb{P}(Z^*=1 \mid \Delta_{Y_0}=1, A=a, \Delta_A=1, L_0=l_0)\mathbb{P}(L_0=l_0) \\
&=& \sum_{l_0}\sum_{y_0}\sum_{l_1}\mathbb{P}(Z^*=1 \mid L_1=l_1, Y_0=y_0, \Delta_{Y_0}=1, A=a, \Delta_A=1, L_0=l_0) \times \\
&& \quad \mathbb{P}(L_1=l_1, Y_0=y_0 \mid \Delta_{Y_0}=1, A=a, \Delta_A=1, L_0=l_0)\mathbb{P}(L_0=l_0)\\
&& \text{by } Z^*=0 \text{ when } Y_0=1 \\ 
&=&  \sum_{l_0}\sum_{l_1}\mathbb{P}(Z^*=1 \mid L_1=l_1, Y_0=0, \Delta_{Y_0}=1, A=a, \Delta_A=1, L_0=l_0) \times \\
&& \quad \mathbb{P}(L_1=l_1, Y_0=0 \mid \Delta_{Y_0}=1, A=a, \Delta_A=1, L_0=l_0)\mathbb{P}(L_0=l_0)\\
&& \text{by } Z^* \perp \Delta_{Y_1} \mid L_1, Y_0=0, \Delta_{Y_0}=1, A=a, \Delta_A=1, L_0 \\ 
&=&  \sum_{l_0}\sum_{l_1}\mathbb{P}(Z^*=1 \mid \Delta_{Y_1}=1, L_1=l_1, Y_0=0, \Delta_{Y_0}=1, A=a, \Delta_A=1, L_0=l_0) \times \\
&& \quad \mathbb{P}(L_1=l_1, Y_0=0 \mid \Delta_{Y_0}=1, A=a, \Delta_A=1, L_0=l_0)\mathbb{P}(L_0=l_0)\\
&=& \mathbb{E}\bigg\{  \mathbb{E}\big[   \mathbb{E}(Y_1 \mid \Delta_{Y_1}=1, L_1, Y_0=0, \Delta_{Y_0}=1, A=a, \Delta_A=1, L_0) \mid \Delta_{Y_0}=1, A=a, \Delta_A=1, L_0  \big] \bigg\}    
\end{eqnarray*}

For the corresponding statistical estimand to be well-defined, we also need the following positivity assumptions: $P(\Delta_{Y_1}=1 \mid L_1, Y_0=0, \Delta_{Y_0}=1, A=a, \Delta_A=1, L_0) > 0$ a.e. in addition to the positivity assumptions for the denominator.

\newpage

\section*{Appendix S4: Alternative time-ordering}

In Figure~3 in the main text, we specified the causal model to reflect the time-ordering in SEARCH-TB, our real-data example.\cite{Marquez2020, marquez2024community,abbott2024incident} Specifically, the pre-exposure covariates $L_0$ exposure $A$ were measured in the parent SEARCH study and before baseline TB status $Y_0$ in the sub-study.  Here, we consider an alternative timeline where the exposure was measured after the baseline outcome, and both are still subject to missingness. The corresponding NPSEM is 
\begin{align*}
&L_0  =f_{L_0}(U_{L_0}) \\
&\Delta_{Y_0} = f_{\Delta_{Y_0}}(L_0, U_{\Delta_{Y_0}}) \\
&Y_0 = f_{Y_0}(L_0, \Delta_{Y_0}, U_{Y_0}) \\
&\Delta_{A} = f_{\Delta_{A}}(L_0, \Delta_{Y_0}, Y_0, U_{\Delta_{A}}) \\
&A = f_{A}(L_0, \Delta_{Y_0}, Y_0, \Delta_{A}, U_{A}) \\
&L_1 = f_{L_1}(L_0, \Delta_{Y_0}, Y_0, \Delta_{A}, A, U_{L_1}) \\
&\Delta_{Y_1} = f_{\Delta_{Y_1}}(L_0, \Delta_{Y_0}, Y_0, \Delta_{A}, A, L_1, U_{\Delta_{Y_1}}) \\
&Y_1 = f_{Y_1}(L_0, \Delta_{Y_0}, Y_0, \Delta_{A}, A, L_1, \Delta_{Y_1}, U_{Y_1}) 
\end{align*}

Let $Y_0^*=Y_0^{\Delta_{Y_0}=1}$ be the counterfactual outcome \emph{at baseline} under a hypothetical intervention to ensure its measurement. Now let  $Y_1^*=Y_1^{\Delta_{Y_0}=1,\Delta_A=1, A=a, \Delta_{Y_1}=1}$ be the counterfactual outcome \emph{at endline} under hypothetical interventions to ensure measurement of the exposures and outcomes and to ``set'' the exposure to level $A=a$. In this scenario, the target estimand is again given by
$$\mathbb{P}(Y_1^*=1 \mid Y_0^*=0) $$
where we emphasize the general form is the same, but the definition of the counterfactuals have changed. Again we are conditioning on the counterfactual outcome at baseline; hence, this is also a Counterfactual Strata Effect.\citep{Balzer2017CascadeMethods,petersen_association_2017,balzer2020far,Balzer2021twostage, nugent2023blurring, petersen_eurosim_2024, gupta_when_2024, havlir_hiv_2019, marquez2024community}

For identification, we again re-express the estimand in terms of the ratio: 
$$\frac{\mathbb{P}(Y_1^*=1, Y_0^*=0)}{\mathbb{P}(Y_0^*=0)}$$

Under the standard assumptions for missing outcomes,  we can identify 1 minus the denominator as
\begin{eqnarray*}
\mathbb{P}(Y_0^* = 1)&=&\sum_{l_0}\mathbb{P}(Y_0^*=1 \mid L_0=l_0)\mathbb{P}(L_0=l_0) \\
&&  Y_0^* \perp \Delta_{Y_0} \mid  L_0\\ 
&=& \sum_{l_0}\mathbb{P}(Y_0^*=1 \mid \Delta_{Y_0}=1, L_0=l_0)\mathbb{P}(L_0=l_0) \\
&=& \mathbb{E}\big[\mathbb{E}(Y_0 \mid \Delta_{Y_0}=1, L_0) \big]
\end{eqnarray*}
along with the corresponding positivity assumption.

\vspace{2em}

For the numerator, again let $Z^*=\mathbb{I}(Y_1^*=1, Y_0^*=0)$. Then under analogous assumptions to Appendix S3.3, we can identify the numerator $\mathbb{P}(Y_1^*=1, Y_0^*=0)=\mathbb{P}(Z^*=1)$.
\begin{eqnarray*}
\mathbb{P}(Z^* = 1)
&=& \sum_{l_0}\sum_{l_1}\mathbb{P}(Y_1=1 \mid \Delta_{Y_1}=1, L_1=l_1,  A=a, \Delta_A=1, Y_0=0, \Delta_{Y_0}=1, L_0=l_0) \times \\
&&\quad \mathbb{P}(L_1=l_1 \mid A=a, \Delta_A=1, Y_0=0, \Delta_{Y_0}=1,  L_0=l_0) \times\\
&& \quad  \mathbb{P}(Y_0=0 \mid \Delta_{Y_0}=1,  L_0=l_0)\mathbb{P}(L_0=l_0) 
\end{eqnarray*}
which can be written in terms of iterated conditional expectations:
\begin{small}
    $$ \mathbb{E}[ \mathbb{E}(   \mathbb{E}\left(   \mathbb{E}(Y_1 \mid \Delta_{Y_1}=1, L_1, A=a,\Delta_A=1, Y_0=0, \Delta_{Y_0}=1, L_0) \mid  A=a, \Delta_A=1, Y_0=0, \Delta_{Y_0}=1, L_0  \right)  \mid  \Delta_{Y_0}=1, L_0)]$$
\end{small}

Putting it all together, our statistical estimand is given by
\begin{footnotesize}
\begin{eqnarray*}
\Psi(\mathbb{P};a) =
    \frac{\mathbb{E}[ \mathbb{E}(   \mathbb{E}\left(   \mathbb{E}(Y_1 \mid \Delta_{Y_1}=1, L_1, A=a,\Delta_A=1, Y_0=0, \Delta_{Y_0}=1, L_0) \mid  A=a, \Delta_A=1, Y_0=0, \Delta_{Y_0}=1, L_0  \right)  \mid  \Delta_{Y_0}=1, L_0)]    } 
    { 1-  \mathbb{E}\big[\mathbb{E}(Y_0 \mid \Delta_{Y_0}=1, L_0) \big]  } 
\end{eqnarray*}
\end{footnotesize}
for exposure level $A=a$.
As before, we can define associations in terms of contrasts $\Psi(\mathbb{P};a)$ at different exposure levels.

\newpage

\section*{Appendix S5: Principal Strata Effects}

We emphasize that Counterfactual Strata Effects are distinct from Principal Strata Effects.\citep{frangakis2002principal,grilli2007university, grilli2008nonparametric,page2015principal} 
For simplicity, suppose that we did not have missingness, and define $Y_0^a$ and $Y_1^a$ as the counterfactual outcome at baseline and endline under exposure $A=a$. 
Then in SEARCH-TB, principal stratification could be applied to define the effect of alcohol on incident TB among the unobservable subset of participants who would have \emph{always} been at-risk of TB at baseline regardless of their alcohol use: $\mathbb{P}(Y_1^1=1 \mid Y_0^1=Y_0^0=0)$ versus $\mathbb{P}(Y_1^0=0 \mid Y_0^1=Y_0^0=0)$. 
Instead, Counterfactual Strata Effects enable us to define the effect of alcohol on incident TB among the \emph{counterfactual} population of persons at-risk at baseline under alcohol use $A=a$:
$\mathbb{P}(Y_1^1=1 \mid Y_0^1=0)$ versus $\mathbb{P}(Y_1^0=1 \mid Y_0^0=0)$. Here, the at-risk population is exposure-specific, and our estimand is contrasting incidence with alcohol use among those who \emph{would be} at-risk with alcohol use versus the incidence without alcohol use among those who \emph{would be} at-risk without alcohol use. Altogether, 
our estimand captures the effect of alcohol use on prevalent TB at baseline and on incident TB by endline. 
Of course, we could also directly examine the effect on prevalent TB at baseline: $\mathbb{P}(Y_0^1=1)$ versus $\mathbb{P}(Y_0^0=1)$. We also reiterate that Counterfactual Strata Effects also arise when the exposure does not impact baseline outcome (Appendix S4) or, more generally, the focus population.\cite{Balzer2017CascadeMethods,petersen_association_2017,balzer2020far,nakato2025measurementmediateseffect}

\newpage

\section*{Appendix S6: The Delta Method for Ratio-Estimands}

The Delta method is commonly applied to derive influence curve-based inference for asymptotically linear estimators.\cite{vanderVaart1998} Recall our exposure-specific estimand for Figure 3 is given by the following ratio: \begin{eqnarray}
\Psi(\mathbb{P};a) =
    \frac{\mathbb{E}[  \mathbb{E}\left(   \mathbb{E}(Y_1 \mid \Delta_{Y_1}=1, L_1, Y_0=0, \Delta_{Y_0}=1, A=a, \Delta_A=1, L_0) \mid \Delta_{Y_0}=1, A=a, \Delta_A=1, L_0  \right) ]    } { 1-  \mathbb{E} \left[\mathbb{E}(Y_0 \mid \Delta_{Y_0}=1, A=a, \Delta_A=1, L_0)\right]  } 
\label{eq:final}
\end{eqnarray}
Following the notation of Moore and van der Laan,\cite{Moore2009} we denote the numerator with $\mu_1$ and the denominator with $\mu_0$.
 Let $IC_1$ and $IC_0$ be the corresponding influence curves. Then based on the Delta method, the influence curve for $log[\Psi(\mathbb{P};a)]$ is given by
$IC(a)=   1/\mu_1\times IC_1 - 1/\mu_0 \times IC_0$.
We obtain a variance estimate with the sample variance of $IC(a)$ scaled by sample size $N$. Using that variance estimate, we conduct hypothesis testing and create 95\% confidence intervals before exponentiating to transform back to the original scale.

\vspace{2em}

Now consider the relative association: 
\begin{eqnarray}
    RR = \frac{\Psi(\mathbb{P};1)}{\Psi(\mathbb{P};0)} = \frac{\mu_1/\mu_0}{\mu_3/\mu_2}
\end{eqnarray}
with corresponding subscripts for the influence curves of the TMLEs. Then the influence curve for the $log(RR)$ is 
given by
\begin{eqnarray}
IC_{RR}= (-1/\mu_0)\times IC_0 + (1/\mu_1)\times IC_1 + (1/\mu_2)\times IC_2 - (1/\mu_3)\times IC_3
\end{eqnarray}

We obtain inference as above.

\newpage

\section*{Appendix S7: Simulation Study}

We conducted a simulation study designed to mimic the causal model given in 
Figure~3 in the main text. The data generating process is as follows where $\text{expit}(x) = (1 + e^{-x})^{-1}$. 
\begin{itemize}
\item Baseline covariates: $L_{0,1} \sim \text{Bernoulli}(0.5)$, $L_{0,2}\sim \text{Uniform}(0,1)$, and $L_{0,3} \sim \text{Uniform}(0,1)$. 
\item Measurement indicator for the exposure: $\Delta_A \sim \text{Bernoulli}(\text{expit}(1.7 + 0.4L_{0,1} - 0.5L_{0,3}))$
\item Exposure: $A \sim \text{Bernoulli}(\text{expit}(-0.3 + 1.5L_{0,1} - 1.2L_{0,2}))$, observed only when $\Delta_A = 1$.
\item Measurement indicator for the baseline outcome: $\Delta_{Y_0} \sim \text{Bernoulli}(\text{expit}(2.0 + 0.3L_{0,1} - 0.4L_{0,2} - 0.3A + 0.2\Delta_A))$
\item Baseline outcome: $Y_0 \sim \text{Bernoulli}(\text{expit}(-2.6 + 1.2L_{0,1} + 0.8L_{0,2} - 0.5L_{0,3} + 0.7A + 0.3A \cdot L_{0,1}))$, observed only when $\Delta_{Y_0} = 1$.
\item Time-varying covariates: $L_{1,1} = \text{clip}(\mathcal{N}(0.3A + 0.2L_{0,1} + 0.15L_{0,2},\ 0.25^2), 0, 1)$ and $L_{1,2} \sim \text{Bernoulli}(\text{expit}(-0.3 + 1.2A + 0.8L_{0,1} - 0.6L_{0,3} + 0.5Y_0))$
\item Measurement indicator for the endline outcome: $\Delta_{Y_1} \sim \text{Bernoulli}(\text{expit}(1.1 + 5.5L_{1,1}^2 A - 5.0L_{0,3}L_{1,1}L_{1,2} - 4.0Y_0 A + 0.5L_{0,1}))$, among those with $\Delta_{Y_0}=1$ and equal to 0 otherwise.
\item Endline outcome: $Y_1 \sim \text{Bernoulli}(\text{expit}(-2.0 + 5.0A \cdot L_{1,1}^2 - 4.5L_{0,3}^2 L_{1,2} + 1.5L_{0,1} + 1.2A))$, observed only when $\Delta_{Y_1}=1$.
\end{itemize}
By generating $10^6$ counterfactuals and taking the  mean, we calculated the true value of the exposure-specific estimands:
\begin{align*}
\Psi^*(\mathbb{P};1) & \ = \ \mathbb{P}(Y_1^1=1 \mid Y_0^1=0) \\
\Psi^*(\mathbb{P};0) & \ = \ \mathbb{P}(Y_1^0=1 \mid Y_0^0=0)
\end{align*}
where $Y_0^a$ and $Y_1^a$ denote the counterfactual outcomes under  $A=a$ and complete measurement and where we have added asterisks to emphasize that these are causal estimands (vs. statistical estimands). We also calculated the relative and absolute effects as $\Psi^*(\mathbb{P};1) \div \Psi^*(\mathbb{P};1)$ and $\Psi^*(\mathbb{P};1) - \Psi^*(\mathbb{P};1)$, respectively.


\vspace{2em}

We implemented TMLE, G-computation, and inverse probability weighting (IPW) for the corresponding statistical estimands (i.e., Equation 2 in the main text). Mirroring the real-data analysis, we used Super Learner within TMLE to combine predictions from generalized linear models, the mean, and multivariate adaptive regression splines with and without covariate screening. We used parametric regressions in the singly robust approaches. Also mirroring the real-data analysis, we implemented  
a ``Na{\"i}ve-adjusted'' approach: subset on participants with complete data ($\Delta_A=\Delta_{Y_0}=\Delta_{Y_1}=1$) and at-risk at baseline ($Y_0=0$) and then implemented TMLE with adjustment for confounding. 
For completeness in the simulations, we also implemented a ``Na{\"i}ve-unadjusted'' approach: 
subset on participants with complete data ($\Delta_A=\Delta_{Y_0}=\Delta_{Y_1}=1$)  and at-risk at baseline ($Y_0=0$) and then take the empirical mean outcome  within exposure groups.
Influence curve-based inference was  used for variance estimation and confidence interval construction. We ran 1000 simulation studies at a sample size of $N=20000$ to reflect the motivating example.  

\subsubsection*{Appendix S7.1: Simulation Results}

The true values of the causal estimands were as follows: the expected outcome under the exposure $\Psi^*(\mathbb{P};1)=0.510$, the expected outcome under no exposure $\Psi^*(\mathbb{P};0)=0.177$, the risk ratio $RR=2.879$, and the risk difference $RD=0.333$.
Table~\ref{tab:sim_20k} provides results in terms of bias (average deviation between the point estimate and truth), empirical standard deviation (standard deviation [SD] of the point estimate across simulation replicates), the ratio of bias to empirical SD, and 95\% confidence interval coverage (proportion of calculated intervals containing the truth). We report the bias-to-SD ratio as a diagnostic for the asymptotic behavior of our estimators: since the standard deviation shrinks at rate $\sqrt{N}$ while systematic bias remains fixed, a non-negligible bias-to-SD ratio indicates performance will deteriorate as sample size grows.

\vspace{2em}

For all estimands,  TMLE with Super Learner achieved  nominal coverage, had negligible  bias, and had bias-to-SD ratios well below 0.2. 
In contrast, G-computation and IPW only achieved nominal coverage for the expected outcome under the exposure $\Psi^*(\mathbb{P};1)$, but not the other estimands. Indeed, the coverage of G-computation was  93.2\% for the expected outcome under no exposure  $\Psi^*(\mathbb{P};0)$, 91.6\% for the risk ratio, and 92.1\% for the risk difference.  
The coverage of IPW  was only  12.3\% for the expected outcome under no exposure  $\Psi^*(\mathbb{P};0)$, 19.7\% for the risk ratio, and 58.9\% for the risk difference.

\begin{table}[H]
\centering
\caption{Over 1000 simulation iterations, estimator performance for TMLE, G-computation (``G-comp''), inverse probability weighting (IPW), and the na{\"i}ve approach with and without confounding adjustment.} 
\label{tab:sim_20k}
\begin{tabular}{llrrrr}
  \toprule
Estimand & Estimator & Bias & Emp SD & Bias/SD & Coverage \\ 
  \midrule
$\Psi^*(\mathbb{P};1)$ & TMLE & 0.0014 & 0.0091 & 0.1560 & 0.9440 \\ 
  $\Psi^*(\mathbb{P};1)$ & G-comp & 0.0053 & 0.0083 & 0.6435 & 0.9500 \\ 
  $\Psi^*(\mathbb{P};1)$ & IPW & -0.0029 & 0.0089 & -0.3231 & 0.9690 \\ 
  $\Psi^*(\mathbb{P};1)$ & Na{\"i}ve-adjusted & 0.0565 & 0.0088 & 6.4347 & 0.0000 \\ 
  $\Psi^*(\mathbb{P};1)$ & Na{\"i}ve-unadjusted & 0.1089 & 0.0080 & 13.5559 & 0.0000 \\ 
   \midrule
$\Psi^*(\mathbb{P};0)$ & TMLE & 0.0003 & 0.0058 & 0.0578 & 0.9530 \\ 
  $\Psi^*(\mathbb{P};0)$ & G-comp & -0.0024 & 0.0054 & -0.4546 & 0.9320 \\ 
  $\Psi^*(\mathbb{P};0)$ & IPW & 0.0179 & 0.0063 & 2.8376 & 0.1230 \\ 
  $\Psi^*(\mathbb{P};0)$ & Na{\"i}ve-adjusted & 0.0102 & 0.0060 & 1.7023 & 0.6290 \\ 
  $\Psi^*(\mathbb{P};0)$ & Na{\"i}ve-unadjusted & -0.0151 & 0.0052 & -2.9191 & 0.0380 \\ 
   \midrule
Risk Ratio & TMLE & 0.0056 & 0.1068 & 0.0525 & 0.9510 \\ 
  Risk Ratio & G-comp & 0.0733 & 0.1012 & 0.7247 & 0.9160 \\ 
  Risk Ratio & IPW & -0.2765 & 0.0953 & -2.9025 & 0.1970 \\ 
  Risk Ratio & Na{\"i}ve-adjusted & 0.1484 & 0.1054 & 1.4071 & 0.7240 \\ 
  Risk Ratio & Na{\"i}ve-unadjusted & 0.9454 & 0.1316 & 7.1833 & 0.0000 \\ 
   \midrule
Risk Difference & TMLE & 0.0011 & 0.0107 & 0.1012 & 0.9460 \\ 
  Risk Difference & G-comp & 0.0078 & 0.0097 & 0.7978 & 0.9210 \\ 
  Risk Difference & IPW & -0.0208 & 0.0109 & -1.9127 & 0.5890 \\ 
  Risk Difference & Na{\"i}ve-adjusted & 0.0463 & 0.0104 & 4.4549 & 0.0100 \\ 
  Risk Difference & Na{\"i}ve-unadjusted & 0.1240 & 0.0095 & 13.0900 & 0.0000 \\ 
   \bottomrule
\end{tabular}
\end{table}

We emphasize that TMLE, G-computation, and IPW rely on the same identification result. Therefore, differences in their performance are due to differences in statistical estimation. 
To explore further, Table~\ref{tab:decomp_20k} provides their performance for the numerator and denominator of our estimands for $a=\{1,0\}$:
\begin{align*}
\Psi^*(\mathbb{P};a) & \ = \ \mathbb{P}(Y_1^1=1 \mid Y_0^1=0)   \ = \ \frac{\mathbb{P}(Y_1^1=1, Y_0^1=0)}{\mathbb{P}(Y_0^1=0)}
\end{align*}
TMLE had negligible bias for all components; its bias-to-SD ratio remained below 0.2. In contrast, G-computation exhibited meaningful bias, as indicated by the bias-to-SD ratio, for all components except for the numerator under $A=1$. 
The numerator/denominator decomposition identifies the source IPW's poor performance: the numerator for $A=0$ is substantially biased with a bias-to-SD ratio of $2.79$. This bias propagates directly into meaningful bias and poor coverage for the expected outcome under no exposure $\Psi^*(\mathbb{P};0)$, the risk ratio, and the risk difference.

\begin{table}[H]
\centering
\caption{Further details on estimator performance with TMLE, G-comp, and IPW for the exposure-specific components of the ratio-based statistical estimands.} 
\label{tab:decomp_20k}
\begin{tabular}{lllrrrrr}
  \toprule
Estimator & Exposure & Component & Truth & Mean Est & Bias & Emp SD & Bias/SD \\ 
  \midrule
TMLE & $A=1$ & Numerator & 0.3593 & 0.3602 & 0.0009 & 0.0071 & 0.1274 \\ 
  TMLE & $A=1$ & Denominator & 0.7047 & 0.7045 & -0.0002 & 0.0056 & -0.0346 \\ 
  TMLE & $A=0$ & Numerator & 0.1499 & 0.1501 & 0.0002 & 0.0050 & 0.0419 \\ 
  TMLE & $A=0$ & Denominator & 0.8464 & 0.8460 & -0.0004 & 0.0044 & -0.0957 \\ 
   \midrule
G-comp & $A=1$ & Numerator & 0.3593 & 0.3604 & 0.0011 & 0.0066 & 0.1686 \\ 
  G-comp & $A=1$ & Denominator & 0.7047 & 0.6996 & -0.0051 & 0.0054 & -0.9529 \\ 
  G-comp & $A=0$ & Numerator & 0.1499 & 0.1473 & -0.0026 & 0.0046 & -0.5549 \\ 
  G-comp & $A=0$ & Denominator & 0.8464 & 0.8435 & -0.0029 & 0.0044 & -0.6663 \\ 
   \midrule
IPW & $A=1$ & Numerator & 0.3593 & 0.3571 & -0.0021 & 0.0069 & -0.3099 \\ 
  IPW & $A=1$ & Denominator & 0.7047 & 0.7045 & -0.0002 & 0.0056 & -0.0344 \\ 
  IPW & $A=0$ & Numerator & 0.1499 & 0.1649 & 0.0151 & 0.0054 & 2.7942 \\ 
  IPW & $A=0$ & Denominator & 0.8464 & 0.8460 & -0.0004 & 0.0045 & -0.0950 \\ 
  \bottomrule
  \end{tabular}
\end{table}

\vspace{2em}

Now we consider the na{\"i}ve approaches, which are targeting a different (incorrect) estimand. Therefore, there was a substantial ``causal gap'', even when adjusting for confounding.\citep{Petersen2014roadmap, dang2023_roadmap, Dang2023, gruber_2024, Nance2024}
By simply subsetting on participants with complete data and baseline risk, the na{\"i}ve approaches failed to appropriately account for differential missingness. As shown in Table~\ref{tab:sim_20k}, the na{\"i}ve approaches exhibited meaningful bias for all estimands. 
The coverage of na{\"i}ve approach with confounding adjustment  was 0\% for the expected outcome under the exposure $\Psi^*(\mathbb{P};1)$, 62.9\% for the expected outcome under no exposure  $\Psi^*(\mathbb{P};0)$, 72.4\% for the risk ratio, and 1\% for the risk difference. 
Finally, the coverage of na{\"i}ve approach without confounding adjustment  was $<$5\% for all estimands.

\vspace{2em}

Altogether, our simulation results support the robustness of TMLE with Super Learner in settings resembling our real-data analysis. Our simulation results further suggest the practical differences between TMLE, G-computation, and IPW in SEARCH-TB are due to TMLE's flexible adjustment for confounding and differential missingness. Again, all three approaches are targeting the same estimand. Therefore, the simulation and applied results highlight how theoretical properties (e.g., double robustness) translate into practical impact.

\newpage

\section*{Appendix S8: Accounting for outcome dependence}

Here, we outline an approach to account for the  dependence of participant outcomes within groups or clusters, such as households, schools, hospitals, or communities. Such dependence could arise due to shared exposures and/or the spread of social behaviors or infectious diseases. In our running example, TB is transmitted from person to person. This dependence should be reflected in the corresponding causal model  (e.g., \citep{halloran1991study, halloran1995causal, hudgens2008toward,balzer_new_2019}).
By following the Causal Roadmap  or a similar framework for causal inference,\citep{Petersen2014roadmap,Hernan2016} we  specify causal models encoding our knowledge about the hierarchical data generating process without imposing parametric modeling assumptions  --- in contrast to more traditional approaches, such as generalizing estimating equations or mixed effects models (e.g., \cite{laan_estimating_2013, balzer_new_2019, Balzer2021twostage, nugent2023blurring, nugent2024causal}). 

\vspace{2em}

Suppose it is reasonable to assume that participant outcomes  are dependent within households, but effectively independent between households. Then our causal model would be specified at the household-level, and identification would consider the influence of other household members as well as community-level factors. (See, for example, \cite{balzer_new_2019}.)
Concretely, this may involve including community indicators in $L_0$ and summary measures of household-level covariates  in $L_0$ and $L_1$. 
The exact form of the causal model and identification result will depend on the application. 
Going forward, we use ``cluster" to refer to any group considered to be the (conditionally) independent  unit.\cite{laan_estimating_2013, nugent2023blurring, nugent2024causal}

\vspace{2em}

If clustering is present, estimation and inference must be adjusted. First, the cross-validation scheme used within Super Learner must respect the independent unit. Concretely, participants in a given cluster are all assigned to the same sample-split. Second, variance estimation must account for clustering. Following Schnitzer et al.,\citep{schnitzer2014effect}  let $m=\{1,\ldots, M\}$ index the clusters and $j=\{1,\ldots, Z_m\}$  index for participants in cluster $m$. Then the total number of participants is $N=\sum_m Z_m$, and  the asymptotic linearity result is re-expressed  as 
$\hat{\Psi} - \Psi 
         = \frac{1}{M}\sum_{m=1}^M \left( \sum_{j \in Z_m} D_{m,j}\frac{M}{N} \right)$
where $D_{m,j}$ denotes the influence curve for the $j^{th}$ participant in the $m^{th}$ cluster and where we suppressed the second-order remainder term for notational convenience.\citep{schnitzer2014effect} Altogether, $X_m = \frac{M}{N} \sum_{j \in Z_m} D_{m,j}$
is the cluster-level influence curve, which has aggregated the individual-level influence curves within  cluster $m$ and is weighted by the ratio of the number of clusters to the number of individuals $M/N$.  We then proceed with variance estimation using the cluster-level influence curve. This approach is equivalent to using an independent working correlation matrix when obtaining robust (sandwich-based) inference. (As detailed \citep{benitez2023defining}, a slightly different formulation is needed if the causal estimand weights clusters, versus individuals, equally.)

\newpage

\section*{Appendix S9: Missingness DAGs}

There is a growing interest in the use of missingness graphs to represent studies with  missingness on multiple variables and to assess whether causal effects can be identified (termed ``recovered'').\citep{mohan2013graphical,mohan2014recovering} In particular, Moreno-Betancur and colleagues introduced complete-data DAGs (c-DAGs), missingness DAGs (m-DAGs), and their link.\cite{moreno2018canonical} They also provided a series of ``canonical'' m-DAGs for causal effects of point-treatment exposures.\cite{moreno2018canonical, zhang2024recoverability} Holovchak et al. extended this work for the effects of longitudinal exposures, while questioning the practical utility of the m-DAG approach.\citep{holovchak2024recoverability} Specifically, they noted, ``no general algorithms are available to decide on recoverability, and decisions have to be made on a case-by-case basis''.\citep{holovchak2024recoverability}
As an alternative, we demonstrated how the Causal Roadmap, an established framework for causal inference, can used to define, identify, and estimate causal estimands, 
including the average treatment effect under missing exposures and Counterfactual Strata Effects.

\clearpage 
\bibliographystyle{unsrtnat}
\bibliography{references}

@article{Gruber2024,
  author  = {Gruber, Susan and Phillips, Richard V. and Lee, Heejung and Ho, Mei and Concato, Joseph and van der Laan, Mark J.},
  title   = {Targeted Learning: Toward a Future Informed by Real-World Evidence},
  journal = {Statistics in Biopharmaceutical Research},
  year    = {2024},
  volume  = {16},
  number  = {1},
  pages   = {11--25},
  doi     = {10.1080/19466315.2023.2182356}
}

@article{Groenwold_missing_2012,
	author = {R.H. Groenwold and I.R. White and A.R Donders and J.R. Carpenter and D.G. Altman and K.G. Moons},
	journal = {CMAJ},
	pages = {1265-1269},
	title = {Missing covariate data in clinical research: when and when not to use the missing-indicator method for analysis},
	volume = {184},
    number= {11},
	year = {2012}}

@article{petersen_association_2017,
	title = {Association of {Implementation} of a {Universal} {Testing} and {Treatment} {Intervention} {With} {HIV} {Diagnosis}, {Receipt} of {Antiretroviral} {Therapy}, and {Viral} {Suppression} in {East} {Africa}},
	volume = {317},
	issn = {1538-3598},
	doi = {10.1001/jama.2017.5705},
	number = {21},
	journal = {JAMA},
	author = {Petersen, Maya and Balzer, Laura and Kwarsiima, Dalsone and Sang, Norton and Chamie, Gabriel and Ayieko, James and Kabami, Jane and Owaraganise, Asiphas and Liegler, Teri and Mwangwa, Florence and Kadede, Kevin and Jain, Vivek and Plenty, Albert and Brown, Lillian and Lavoy, Geoff and Schwab, Joshua and Black, Douglas and van der Laan, Mark and Bukusi, Elizabeth A. and Cohen, Craig R. and Clark, Tamara D. and Charlebois, Edwin and Kamya, Moses and Havlir, Diane},
	month = jun,
	year = {2017},
	pmid = {28586888},
	pmcid = {PMC5734234},
	pages = {2196--2206},
}

@article{Moore2009,
	author = {Moore, K.L. and van der Laan, M.J.},
	doi = {10.1002/sim.3445},
	journal = {Statistics in Medicine},
	number = {1},
	pages = {39--64},
	title = {Covariate Adjustment in Randomized Trials with Binary Outcomes: {T}argeted Maximum Likelihood Estimation},
	volume = {28},
	year = {2009}}

@article{Wong2021,
	author = {A. Wong and L.B. Balzer},
	journal = {Epidemiology},
	number = {2},
	pages = {228-236},
	title = {State-Level Masking Mandates and {COVID-19} Outcomes in the {U}nited {S}tates: A Demonstration of the Causal Roadmap},
	volume = {33},
	year = {2022}}

@article{Robins2000,
	author = {J.M. Robins and M.A. Hern\'{a}n and B. Brumback},
	journal = {Epidemiology},
	number = {5},
	pages = {550--560},
	title = {Marginal structural models and causal inference in epidemiology},
	volume = {11},
	year = {2000}}

@article{Taubman2009,
	author = {S.L. Taubman and J.M. Robins and M.A. Mittleman and M.A. Hern\'{a}n},
	journal = {International Journal of Epidemiology},
	number = {6},
	pages = {1599--1611},
	title = {Intervening on risk factors for coronary heart disease: an application of the parametric {G-formula}},
	volume = {38},
	year = {2009}}

@book{PearlCausality,
	address = {New York},
	author = {Pearl, J.},
	edition = {2nd},
	publisher = {Cambridge University Press},
	title = {{Causality: Models, Reasoning and Inference}},
	year = {2009}}

@article{dang2023_roadmap,
	author = {L.E. Dang and S. Gruber and H. Lee and I.J. Dahabreh and E.A. Stuart and others},
	journal = {J Clin Transl Sci},
	number = {1},
	pages = {e213},
	title = {A causal roadmap for generating high-quality real-world evidence},
	volume = {7},
	year = {2023}}

@book{whatif,
	address = {Boca Raton},
	author = {M.A. Hern\'{a}n and J.M. Robins},
	publisher = {Chapman \& Hall/CRC},
	title = {Causal Inference: What If},
	url = {https://miguelhernan.org/whatifbook},
	year = {2020}}

@techreport{nakato2025measurementmediateseffect,
      title={A causal framework for evaluating the total effect of strategies
aiming to expand screening and to improve outcomes}, 
      author={Joy Zora Nakato and Janice Litunya and Brian Beesiga and Jane Kabami and James Ayieko and Moses R. Kamya and Gabriel Chamie and Laura B. Balzer},
      year={2026},
      eprint={2506.06267v3},
      archivePrefix={arXiv},
      primaryClass={stat.ME},
      url={https://arxiv.org/abs/2506.06267v3}, 
}

@article{gruber_2024,
author = {Susan Gruber and Rachael V. Phillips and Hana Lee and Martin Ho and John Concato and Mark J. van der Laan and},
title = {Targeted Learning: Toward a Future Informed by Real-World Evidence},
journal = {Statistics in Biopharmaceutical Research},
volume = {16},
number = {1},
pages = {11--25},
year = {2024},
publisher = {ASA Website},
doi = {10.1080/19466315.2023.2182356},
}

@article{lash_bias,
    author = {Lash, Timothy L and Fox, Matthew P and MacLehose, Richard F and Maldonado, George and McCandless, Lawrence C and Greenland, Sander},
    title = {Good practices for quantitative bias analysis},
    journal = {International Journal of Epidemiology},
    volume = {43},
    number = {6},
    pages = {1969-1985},
    year = {2014},
    month = {07},
    issn = {0300-5771},
    doi = {10.1093/ije/dyu149},
    url = {https://doi.org/10.1093/ije/dyu149},
    eprint = {https://academic.oup.com/ije/article-pdf/43/6/1969/2083090/dyu149.pdf},
}

@article{Dang2023,
	author = {L.E. Dang and L.B. Balzer},
	journal = {Epidemiology},
	number = {5},
	pages = {619-623},
	title = {Start with the target trial protocol; then follow the {R}oadmap for causal inference},
	volume = {34},
	year = {2023}}

@book{vanderVaart1998,
	address = {New York},
	author = {A.W. van der Vaart},
	publisher = {Cambridge University Press},
	title = {Asymptotic Statistics},
	year = {1998}}

@article{Hernan2016,
	author = {M.A. Hern\'{a}n and J.M. Robins},
	journal = {American Journal of Epidemiology},
	number = {8},
	pages = {758-764},
	title = {Using Big Data to Emulate a Target Trial When a Randomized Trial Is Not Available},
	volume = {183},
	year = {2016}}

@article{Nance2024,
	author = {N. Nance and M. Petersen and M. {van der Laan} and L.B. Balzer},
	journal = {Epidemiology},
	number = {6},
	pages = {791-800},
	title = {The Causal Roadmap and Simulations to Improve the Rigor and Reproducibility of Real-Data Applications},
	volume = {35},
	year = {2024}}

@article{balzer_new_2019,
	title = {A new approach to hierarchical data analysis: {Targeted} maximum likelihood estimation for the causal effect of a cluster-level exposure},
	volume = {28},
	issn = {0962-2802},
	shorttitle = {A new approach to hierarchical data analysis},
	url = {https://doi.org/10.1177/0962280218774936},
	doi = {10.1177/0962280218774936},
	number = {6},
	journal = {Stat Methods Med Res},
	author = {Balzer, Laura B and Zheng, Wenjing and van der Laan, Mark J and Petersen, Maya L},
	month = jun,
	year = {2019},
	pages = {1761--1780},
}

@article{laan_estimating_2013,
	title = {Estimating the {Effect} of a {Community}-{Based} {Intervention} with {Two} {Communities}},
	volume = {1},
	issn = {2193-3685},
	url = {http://www.degruyter.com/document/doi/10.1515/jci-2012-0011/html},
	number = {1},
	journal = {Journal of Causal Inference},
	author = {van der Laan, Mark J.  and Petersen, Maya and Zheng, Wenjing},
	month = may,
	year = {2013},
	pages = {83--106},
}

@article{Petersen2014roadmap,
	Author = {M.L. Petersen and M.J. van der Laan},
	Journal = {Epidemiology},
	Number = {3},
	Pages = {418-426},
	Title = {Causal Models and Learning from Data: Integrating Causal Modeling and Statistical Estimation},
	Volume = {25},
	Year = {2014}}

@techreport{Balzer2017CascadeMethods,
	author = {L.B. Balzer and J. Schwab and M.J. van der Laan and M.L. Petersen},
	institution = {University of California at Berkeley},
	number = {357},
	title = {Evaluation of Progress Towards the {UNAIDS} 90-90-90 {HIV} Care Cascade: A Description of Statistical Methods Used in an Interim Analysis of the Intervention Communities in the {SEARCH} Study},
	url = {http://biostats.bepress.com/ucbbiostat/paper357/},
	year = {2017}}

@techreport{Balzer2018SAP,
	author = {L.B. Balzer and D.V. Havlir and J. Schwab and M.J. van der Laan and M.L. Petersen and {the SEARCH Collaboration}},
	institution = {arXiv: https://arxiv.org/abs/1808.03231},
	title = {Statistical Analysis Plan for {SEARCH} {}Phase {I}: {H}ealth Outcomes among Adults},
	year = {2018}}

@article{Bang&Robins05,
	author = {H. Bang and J.M. Robins},
	journal = {Biometrics},
	pages = {962--972},
	title = {Doubly robust estimation in missing data and causal inference models},
	volume = {61},
	year = {2005}}

@article{vanderLaan2012towertmle,
	author = {M.J. van der Laan and S. Gruber},
	journal = {The International Journal of Biostatistics},
	number = {1},
	title = {Targeted minimum loss based estimation of causal effects of multiple time point interventions},
	volume = {8},
	year = {2012}}

@article{Marquez2020,
	author = {C. Marquez and M. Atukunda and L.B. Balzer and G. Chamie and others},
	journal = {PloS ONE},
	number = {1},
	pages = {e0228102},
	title = {The Age-Specific Burden and Household and School-Based Predictors of Child and Adolescent Tuberculosis Infection in Rural Uganda},
	volume = {15},
	year = {2020}}

@article{chamie_hybrid_2016,
	author = {Chamie, Gabriel and Clark, Tamara D and Kabami, Jane and Kadede, Kevin and Ssemmondo, Emmanuel and Steinfeld, Rachel and Lavoy, Geoff and Kwarisiima, Dalsone and Sang, Norton and Jain, Vivek and Thirumurthy, Harsha and Liegler, Teri and Balzer, Laura B and Petersen, Maya L and Cohen, Craig R and Bukusi, Elizabeth A and Kamya, Moses R and Havlir, Diane V and Charlebois, Edwin D},
	doi = {10.1016/S2352-3018(15)00251-9},
	issn = {2352-3018},
	journal = {The Lancet HIV},
	language = {en},
	number = {3},
	pages = {e111--e119},
	shorttitle = {A hybrid mobile approach for population-wide {HIV} testing in rural east {Africa}},
	title = {A hybrid mobile approach for population-wide {HIV} testing in rural east {Africa}: an observational study},
	volume = {3},
	year = {2016}}

@article{havlir_hiv_2019,
	author = {Havlir, Diane V. and Balzer, Laura B. and Charlebois, Edwin D. and Clark, Tamara D. and Kwarisiima, Dalsone and Ayieko, James and Kabami, Jane and Sang, Norton and Liegler, Teri and Chamie, Gabriel and {et al.}},
	doi = {10.1056/NEJMoa1809866},
	issn = {0028-4793},
	journal = {New England Journal of Medicine},
	number = {3},
	pages = {219--229},
	title = {{HIV} {Testing} and {Treatment} with the {Use} of a {Community} {Health} {Approach} in {Rural} {Africa}},
	url = {http://www.nejm.org/doi/10.1056/NEJMoa1809866},
	volume = {381},
	year = {2019}}

@conference{petersen_eurosim_2024,
	author = {Petersen, Maya},
address = {Copenhagen, Denmark},
	booktitle = {European Causal Inference Meeting},
	title = {The {C}ausal {R}oadmap in the age of {AI}: from all wheel drive to formula 1},
	year = {2024}
}

@article{robins2008estimation,
  title={Estimation and extrapolation of optimal treatment and testing strategies},
  author={Robins, James and Orellana, Liliana and Rotnitzky, Andrea},
  journal={Statistics in medicine},
  volume={27},
  number={23},
  pages={4678--4721},
  year={2008},
  publisher={Wiley Online Library}
}

@article{benitez2023defining,
  title={Defining and estimating effects in cluster randomized trials: a methods comparison},
  author={Benitez, Alejandra and Petersen, Maya L and van der Laan, Mark J and Santos, Nicole and Butrick, Elizabeth and Walker, Dilys and Ghosh, Rakesh and Otieno, Phelgona and Waiswa, Peter and Balzer, Laura B},
  journal={Statistics in medicine},
  volume={42},
  number={19},
  pages={3443--3466},
  year={2023},
  publisher={Wiley Online Library}
}

@article{van2007causal,
  title={Causal effect models for realistic individualized treatment and intention to treat rules},
  author={Van der Laan, Mark J and Petersen, Maya L},
  journal={The international journal of biostatistics},
  volume={3},
  number={1},
  year={2007},
  publisher={De Gruyter}
}

@article{hernan2006comparison,
  title={Comparison of dynamic treatment regimes via inverse probability weighting},
  author={Hern{\'a}n, Miguel A and Lanoy, Emilie and Costagliola, Dominique and Robins, James M},
  journal={Basic \& clinical pharmacology \& toxicology},
  volume={98},
  number={3},
  pages={237--242},
  year={2006},
  publisher={Wiley Online Library}
}

@techreport{gupta_when_2024,
	title = {When exposure affects subgroup membership: {Framing} relevant causal questions in perinatal epidemiology and beyond},
	shorttitle = {When exposure affects subgroup membership},
	url = {http://arxiv.org/abs/2401.11368},
	doi = {10.48550/arXiv.2401.11368},
	urldate = {2024-07-11},
	publisher = {arXiv},
	author = {Gupta, Shalika and Balzer, Laura B. and Kamya, Moses R. and Havlir, Diane V. and Petersen, Maya L.},
	month = jan,
	year = {2024},
	note = {arXiv:2401.11368 [stat]},
}

@book{MarkRobins2003,
	Address = {New York Berlin Heidelberg},
	Author = {van der Laan, M.J. and Robins, J.M.},
	Publisher = {Springer-Verlag},
	Title = {{Unified Methods for Censored Longitudinal Data and Causality}},
	Year = {2003}}

@article{horvitzGeneralizationSamplingReplacement1952,
  title = {A {{Generalization}} of {{Sampling Without Replacement From}} a {{Finite Universe}}},
  author = {Horvitz, D. G. and Thompson, D. J.},
  year = {1952},
  journal = {Journal of the American Statistical Association},
  volume = {47},
  number = {260},
  pages = {663--685},
  issn = {0162-1459}
}

@article{frangakis2002principal,
  title={Principal stratification in causal inference},
  author={Frangakis, Constantine E and Rubin, Donald B},
  journal={Biometrics},
  volume={58},
  number={1},
  pages={21--29},
  year={2002},
  publisher={Wiley Online Library}
}

@article{balzer2020far,
  title={Far from {MCAR}: obtaining population-level estimates of {HIV} viral suppression},
  author={Balzer, Laura B and Ayieko, James and Kwarisiima, Dalsone and Chamie, Gabriel and Charlebois, Edwin D and Schwab, Joshua and van der Laan, Mark J and Kamya, Moses R and Havlir, Diane V and Petersen, Maya L},
  journal={Epidemiology (Cambridge, Mass.)},
  volume={31},
  number={5},
  pages={620},
  year={2020},
  publisher={NIH Public Access}
}

@article{rubin1976inference,
  title={Inference and missing data},
  author={Rubin, Donald B},
  journal={Biometrika},
  volume={63},
  number={3},
  pages={581--592},
  year={1976},
  publisher={Oxford University Press}
}

@article{carpenito2022misl,
  title={{MISL}: Multiple imputation by super learning},
  author={Carpenito, Thomas and Manjourides, Justin},
  journal={Statistical Methods in Medical Research},
  volume={31},
  number={10},
  pages={1904--1915},
  year={2022},
  publisher={SAGE Publications}
}

@article{laqueur2022supermice,
  title={{SuperMICE}: An ensemble machine learning approach to multiple imputation by chained equations},
  author={Laqueur, Hannah S and Shev, Aaron B and Kagawa, Rose MC},
  journal={American Journal of Epidemiology},
  volume={191},
  number={3},
  pages={516--525},
  year={2022},
  publisher={Oxford University Press}
}

@article{kennedy2020efficient,
  title={Efficient nonparametric causal inference with missing exposure information},
  author={Kennedy, Edward H},
  journal={The International Journal of Biostatistics},
  volume={16},
  number={1},
  pages={20190087},
  year={2020},
  publisher={De Gruyter}
}

@article{zhang2016causal,
  title={Causal inference with missing exposure information: Methods and applications to an obstetric study},
  author={Zhang, Zhiwei and Liu, Wei and Zhang, Bo and Tang, Li and Zhang, Jun},
  journal={Statistical Methods in Medical Research},
  volume={25},
  number={5},
  pages={2053--2066},
  year={2016},
  publisher={SAGE Publications}
}

@article{grilli2008nonparametric,
  title={Nonparametric bounds on the causal effect of university studies on job opportunities using principal stratification},
  author={Grilli, Leonardo and Mealli, Fabrizia},
  journal={Journal of Educational and Behavioral Statistics},
  volume={33},
  number={1},
  pages={111--130},
  year={2008},
  publisher={SAGE Publications}
}

@article{grilli2007university,
  title={University studies and employment: An application of the principal strata approach to causal analysis},
  author={Grilli, Leonardo and Mealli, Fabrizia},
  journal={Effectiveness of University Education in Italy: Employability, Competences, Human Capital},
  pages={219--231},
  year={2007},
  publisher={Springer}
}

@article{nugent2023blurring,
  title={Blurring cluster randomized trials and observational studies: Two-Stage {TMLE} for subsampling, missingness, and few independent units},
  author={Nugent, Joshua R and Marquez, Carina and Charlebois, Edwin D and Abbott, Rachel and Balzer, Laura B and SEARCH Collaboration},
  journal={Biostatistics},
  volume={24},
  pages={599--616},
  year={2024},
  publisher={Oxford University Press}
}

@article{cornelisz2020addressing,
  title={Addressing missing data in randomized clinical trials: A causal inference perspective},
  author={Cornelisz, Ilja and Cuijpers, Pim and Donker, Tara and van Klaveren, Chris},
  journal={{PloS One}},
  volume={15},
  number={7},
  pages={e0234349},
  year={2020},
  publisher={Public Library of Science San Francisco, CA USA}
}

@article{sterne2009multiple,
  title={Multiple imputation for missing data in epidemiological and clinical research: potential and pitfalls},
  author={Sterne, Jonathan AC and White, Ian R and Carlin, John B and Spratt, Michael and Royston, Patrick and Kenward, Michael G and Wood, Angela M and Carpenter, James R},
  journal={{BMJ}},
  volume={338},
  year={2009},
  publisher={British Medical Journal Publishing Group}
}

@article{wells2013strategies,
  title={Strategies for handling missing data in electronic health record derived data},
  author={Wells, Brian J and Chagin, Kevin M and Nowacki, Amy S and Kattan, Michael W},
  journal={Egems},
  volume={1},
  number={3},
  year={2013},
  publisher={Ubiquity Press}
}

@article{little2012prevention,
  title={The prevention and treatment of missing data in clinical trials},
  author={Little, Roderick J and D'Agostino, Ralph and Cohen, Michael L and Dickersin, Kay and Emerson, Scott S and Farrar, John T and Frangakis, Constantine and Hogan, Joseph W and Molenberghs, Geert and Murphy, Susan A and others},
  journal={New England Journal of Medicine},
  volume={367},
  number={14},
  pages={1355--1360},
  year={2012},
  publisher={Mass Medical Soc}
}

@article{cole2023missing,
  title={Missing outcome data in epidemiologic studies},
  author={Cole, Stephen R and Zivich, Paul N and Edwards, Jessie K and Ross, Rachael K and Shook-Sa, Bonnie E and Price, Joan T and Stringer, Jeffrey SA},
  journal={American Journal of Epidemiology},
  volume={192},
  number={1},
  pages={6--10},
  year={2023},
  publisher={Oxford University Press}
}

@article{Balzer2021twostage,
	author = {L.B. Balzer and M. {van der Laan} and J. Ayieko and M. Kamya and others},
journal = {Biostatistics},
	number = {2},
	pages = {502-517},
	title = {Two-Stage {TMLE} to Reduce Bias and Improve Efficiency in Cluster Randomized Trials},
	volume = {24},
	year = {2023}}

@article{schnitzer2014effect,
  title={Effect of breastfeeding on gastrointestinal infection in infants: A targeted maximum likelihood approach for clustered longitudinal data},
  author={Schnitzer, Mireille E and van der Laan, Mark J and Moodie, Erica EM and Platt, Robert W},
  journal={The Annals of Applied Statistics},
  volume={8},
  number={2},
  pages={703},
  year={2014},
  publisher={NIH Public Access}
}

@article{ghazaleh2021handling,
  title={Handling missing data when estimating causal effects with targeted maximum likelihood estimation},
  author={Ghazaleh Dashti, S and Lee, Katherine J and Simpson, Julie A and White, Ian R and Carlin, John B and Moreno-Betancur, Margarita},
  journal={American Journal of Epidemiology},
  volume={193},
  number={7},
  pages={1019--1030},
  year={2024},
  publisher={Oxford University Press}
}

@article{moreno2018canonical,
  title={Canonical causal diagrams to guide the treatment of missing data in epidemiologic studies},
  author={Moreno-Betancur, Margarita and Lee, Katherine J and Leacy, Finbarr P and White, Ian R and Simpson, Julie A and Carlin, John B},
  journal={American Journal of Epidemiology},
  volume={187},
  number={12},
  pages={2705--2715},
  year={2018},
  publisher={Oxford University Press}
}

@article{abbott2024incident,
  title={Incident tuberculosis infection is associated with alcohol use in adults in rural {Uganda}},
  author={Abbott, Rachel and Landsiedel, Kirsten and Atukunda, Mucunguzi and Puryear, Sarah B and Chamie, Gabriel and Hahn, Judith A and Mwangwa, Florence and Kakande, Elijah and Petersen, Maya L and Havlir, Diane V and others},
  journal={Clinical Infectious Diseases},
  volume={78},
  pages={ciae304},
  year={2024},
  publisher={Oxford University Press}
}

@book{van2011targeted,
  title={Targeted learning: Causal inference for observational and experimental data},
  author={van der Laan, Mark J and Rose, Sherri and others},
  year={2011},
  publisher={Springer}
}

@article{rose2011targeted,
  title={A targeted maximum likelihood estimator for two-stage designs},
  author={Rose, Sherri and van der Laan, Mark J},
  journal={The international journal of biostatistics},
  volume={7},
  number={1},
  pages={0000102202155746791217},
  year={2011},
  publisher={De Gruyter}
}

@book{van2018targeted,
  title={Targeted {L}earning in data science},
  author={van der Laan, Mark J and Rose, Sherri},
  year={2018},
  publisher={Springer}
}

@article{marquez2024community,
  title={Community-wide universal human immunodeficiency virus ({HIV}) test and treat intervention reduces tuberculosis transmission in rural {Uganda}: A cluster-randomized trial},
  author={Marquez, Carina and Atukunda, Mucunguzi and Nugent, Joshua and Charlebois, Edwin D and Chamie, Gabriel and Mwangwa, Florence and Ssemmondo, Emmanuel and Kironde, Joel and Kabami, Jane and Owaraganise, Asiphas and others},
  journal={Clinical Infectious Diseases},
  volume={78},
  pages={ciad776},
  year={2024},
  publisher={Oxford University Press US}
}

@article{van2007super,
  title={Super learner},
  author={van der Laan, Mark J and Polley, Eric C and Hubbard, Alan E},
  journal={Statistical Applications in Genetics and Molecular Biology},
  volume={6},
  number={1},
  year={2007},
  publisher={De Gruyter}
}

@book{Rubin1987,
  author    = {Donald B. Rubin},
  title     = {Multiple Imputation for Nonresponse in Surveys},
  year      = {1987},
  publisher = {John Wiley \& Sons},
  address   = {New York},
  series    = {Wiley Series in Probability and Statistics},
  isbn      = {9780471087052},
  doi       = {10.1002/9780470316696}
}

@techreport{balzer2017evaluation,
title = {Evaluation of Progress Towards the {UNAIDS} 90-90-90 {HIV} Care Cascade: A Description of Statistical Methods Used in an Interim Analysis of the Intervention Communities in the {SEARCH} Study},
  author={Balzer, Laura and Schwab, Joshua and van der Laan, Mark J and Petersen, Maya L},
  year={2017},
 	institution = {University of California at Berkeley},
	number = {357},
	url = {http://biostats.bepress.com/ucbbiostat/paper357/}
}

@article{juul2024missing,
  title={Missing outcome data in randomised clinical trials of psychological interventions: a review of published trial reports in major psychiatry journals},
  author={Juul, Sophie and Faltermeier, Pascal and Petersen, Johanne Juul and Olsen, Markus Harboe and Andersen, Rebecca Kjaer and Kamp, Caroline Barkholt and Siddiqui, Faiza and Simonsen, Sebastian and Mbuagbaw, Lawrence and Thabane, Lehana and others},
  journal={BMC psychiatry},
  volume={24},
  number={1},
  pages={798},
  year={2024},
  publisher={Springer}
}

@article{medcalf2024addressing,
  title={Addressing missing outcome data in randomised controlled trials: a methodological scoping review},
  author={Medcalf, Ellie and Turner, Robin M and Espinoza, David and He, Vicky and Bell, Katy JL},
  journal={Contemporary clinical trials},
  pages={107602},
  year={2024},
  publisher={Elsevier}
}

@article{robins1994estimation,
  title={Estimation of regression coefficients when some regressors are not always observed},
  author={Robins, James M and Rotnitzky, Andrea and Zhao, Liangping},
  journal={Journal of the American Statistical Association},
  volume={89},
  number={427},
  pages={846--866},
  year={1994},
  publisher={Taylor & Francis}
}

@article{robins1986new,
  title={A new approach to causal inference in mortality studies with a sustained exposure period—Application to control of the healthy worker survivor effect},
  author={Robins, James M.},
  journal={Mathematical Modelling},
  volume={7},
  number={9},
  pages={1393--1512},
  year={1986},
  publisher={Elsevier}
}

@article{young2020causal,
  title={A causal framework for classical statistical estimands in failure-time settings with competing events},
  author={Young, Jessica G and Stensrud, Mats J and Tchetgen Tchetgen, Eric J and Hern{\'a}n, Miguel A},
  journal={Statistics in medicine},
  volume={39},
  number={8},
  pages={1199--1236},
  year={2020},
  publisher={Wiley Online Library}
}

@article{page2015principal,
  title={Principal stratification: A tool for understanding variation in program effects across endogenous subgroups},
  author={Page, Lindsay C and Feller, Avi and Grindal, Todd and Miratrix, Luke and Somers, Marie-Andree},
  journal={American Journal of Evaluation},
  volume={36},
  number={4},
  pages={514--531},
  year={2015},
  publisher={Sage Publications Sage CA: Los Angeles, CA}
}

@book{RothmanModern,
	address = {Phildelphia},
	author = {Rothman, K.J. and Greenland, S. and Lash, T.L.},
	edition = {3rd},
	publisher = {Lippincott Williams \& Wilkins},
	title = {Modern Epidemiology},
	year = {2008}}

@article{nugent2024causal,
  title={Causal Inference in Randomized Trials with Partial Clustering and Imbalanced Dependence Structures},
  author={Nugent, Joshua R and Kakande, Elijah and Chamie, Gabriel and Kabami, Jane and Owaraganise, Asiphas and Havlir, Diane V and Kamya, Moses and Balzer, Laura B},
  journal={arXiv preprint arXiv:2406.04505},
  year={2024}
}

@article{hudgens2008toward,
  title={Toward causal inference with interference},
  author={Hudgens, Michael G and Halloran, M Elizabeth},
  journal={Journal of the american statistical association},
  volume={103},
  number={482},
  pages={832--842},
  year={2008},
  publisher={Taylor \& Francis}
}

@article{halloran1995causal,
  title={Causal inference in infectious diseases},
  author={Halloran, M Elizabeth and Struchiner, Claudio J},
  journal={Epidemiology},
  pages={142--151},
  year={1995},
  publisher={JSTOR}
}

@article{halloran1991study,
  title={Study designs for dependent happenings},
  author={Halloran, M Elizabeth and Struchiner, Claudio J},
  journal={Epidemiology},
  volume={2},
  number={5},
  pages={331--338},
  year={1991},
  publisher={LWW}
}

@inproceedings{mohan2014recovering,
  title={Recovering Probabilistic Queries from Missing Data},
  author={Mohan, Karthika and Pearl, Judea},
  booktitle={Advances in Neural Information Processing Systems},
  year={2014},
  pages={190--198}
}

@article{mohan2013graphical,
  title={Graphical models for inference with missing data},
  author={Mohan, Karthika and Pearl, Judea and Tian, Jin},
  journal={Advances in neural information processing systems},
  volume={26},
  year={2013}
}

@article{holovchak2024recoverability,
  title={Recoverability of Causal Effects in a Longitudinal Study Under Presence of Missing Data},
  author={Holovchak, A and McIlleron, H and Denti, P and Schomaker, M},
  journal={Biostatistics},
  year={2024}
}

@article{zhang2024recoverability,
  title={Recoverability and estimation of causal effects under typical multivariable missingness mechanisms},
  author={Zhang, Jiaxin and Dashti, S Ghazaleh and Carlin, John B and Lee, Katherine J and Moreno-Betancur, Margarita},
  journal={Biometrical Journal},
  volume={66},
  number={3},
  pages={2200326},
  year={2024},
  publisher={Wiley Online Library}
}

@article{ross2020complete,
  title={When is a complete-case approach to missing data valid? The importance of effect-measure modification},
  author={Ross, Rachael K and Breskin, Alexander and Westreich, Daniel},
  journal={American journal of epidemiology},
  volume={189},
  number={12},
  pages={1583--1589},
  year={2020},
  publisher={Oxford University Press}
}

@techreport{mathur2024resurrecting,
  title={Resurrecting complete-case analysis: A defense},
  author={Mathur, Maya B and Shpitser, Ilya and VanderWeele, Tyler J},
  year={2024},
  institution={Center for Open Science}
}

@article{williamson2024assessing,
  title={Assessing treatment effects in observational data with missing confounders: A comparative study of practical doubly-robust and traditional missing data methods},
  author={Williamson, Brian D and Krakauer, Chloe and Johnson, Eric and Gruber, Susan and Shepherd, Bryan E and van der Laan, Mark J and Lumley, Thomas and Lee, Hana and Munoz, Jose J Hernandez and Zhao, Fengyu and others},
  journal={arXiv preprint arXiv:2412.15012},
  year={2024}
}

@techreport{qiu2026efficient,
  title={Efficient Targeted Maximum Likelihood Estimators for Two-Phase Design Problems},
  author={Qiu, Sky and Gruber, Susan and Shaw, Pamela A. and Williamson, Brian D. and van der Laan, Mark J.},
  year={2026},
  eprint={2602.24131},
  archivePrefix={arXiv},
  primaryClass={stat.ME},
  url={https://arxiv.org/abs/2602.24131}
}

@techreport{xu2021missing,
  title={The Missing Covariate Indicator Method is Nearly Valid Almost Always},
  author={Xu, Gang and Song, Mingyang and Zhou, Xin and Wu, Yilun and Pazaris, Mathew and Spiegelman, Donna},
  year={2021},
  eprint={2111.00138},
  archivePrefix={arXiv},
  primaryClass={stat.AP},
  url={https://arxiv.org/abs/2111.00138}
}

@article{tompsett2023target,
  title={Target trial emulation and bias through missing eligibility data: an application to a study of palivizumab for the prevention of hospitalization due to infant respiratory illness},
  author={Tompsett, Daniel and Zylbersztejn, Ania and Hardelid, Pia and De Stavola, Bianca},
  journal={American Journal of Epidemiology},
  volume={192},
  number={4},
  pages={600--611},
  year={2023},
  publisher={Oxford University Press}
}

\end{document}